\documentclass[12pt,reqno]{amsart}
\usepackage{graphicx}
\usepackage{amssymb,amscd,amsbsy}
\usepackage{amsthm}
\newcommand{\noi}{\noindent}

\newcommand{\Z}{{\mathbb Z}}
\newcommand{\C}{{\mathbb C}}
\newcommand{\RB}{{\mathbb R}}

\newcommand\Dg{\mathfrak D}

\def\F{\Phi}

\def\[{\left[}
\def\]{\right]}
\def\({\left(}
\def\){\right)}
\def\t{\theta}
\def\l{\lambda}

\def\lt{\frac{\lambda}{2}}
\def\psib{\overline{\psi}}

\def\d{\delta}
\def\ea{\epsilon_a}
\def\epm{\epsilon_{\pm}}

\def\mi{\mu_{\infty}(\lambda)}
\def\pin{p_{\infty} (\lambda)}
\def\fr{\f_{\rightarrow}}
\def\fl{\f_{\leftarrow}}
\def\Gin{\Gamma_\infty}
\def\G{\Gamma}

\newcommand{\e}{{\boldsymbol e}}
\newcommand{\f}{{\boldsymbol f}}
\newcommand{\jo}{{\boldsymbol j}}

\newcommand{\g}{\gamma}

\newcommand{\eeq}{\end{equation}}
\newcommand{\beq}{\begin{equation}}
\newcommand{\bay}{\begin{eqnarray}}
\newcommand{\ey}{\end{eqnarray}}
\newcommand{\bey}{\begin{eqnarray*}}
\newcommand{\eey}{\end{eqnarray*}}

\newcommand{\R}{\operatorname{res}}

\newcommand{\tr}{\operatorname{trace}}

\newtheorem{thm}{\hspace{\parindent}Theorem}[section]

\newtheorem{lem}[thm]{\hspace{\parindent}Lemma}

\theoremstyle{remark}

\newtheorem*{rem*}{Remark}

\begin{document}

\newcommand{\vse}{\vspace{.2in}}
\numberwithin{equation}{section}

\title{\bf Symplectic structures for  the cubic Schr\"{o}dinger equation in the  
 periodic and scattering case.}
\author{K.L.  Vaninsky}
\thanks{ 35Q53; 58B99. The work is partially supported by NSF grant DMS-9971834.}

\begin{abstract}
We develop a unified approach for construction of symplectic forms for 1D integrable 
equations with the periodic and rapidly decaying initial data. As an example we consider 
the cubic nonlinear Schr\"{o}dinger equation.  
\end{abstract}
\maketitle
\setcounter{section}{0}
\setcounter{equation}{0}
\section{\bf Introduction}

\subsection{General remarks.} The main technical tool for the study  of  soliton systems  is   commutator formalism. 
All fashionable   soliton systems like the  Korteveg--de--Vriez equation (KdV), the cubic nonlinear  Schr\"{o}dinger 
equation (NLS), the $\sin$--Gordon equation,  the Toda lattice,  {\it etc}; have  such representation. Within the commutator formalism   
approach the dynamical system appears as a compatibility  condition for an over-determined system of equations. 
As an  example, we consider   the NLS equation with  repulsive nonlinearity 
\footnote{Prime $'$  signifies derivative in   variable $x$ and dot $\bullet$ in time.} 
$$
i \psi^{\bullet}= -\psi'' + 2 |\psi |^2 \psi,  
$$
where $\psi(x,t)$ is a complex function of spatial variable  $x$ and time $t$. 
The   flow is a compatibility condition for the commutator  
$$
[\partial_t- V_3,  \partial_x-V_2]=0,
$$
with 
$$
V_2(x,t)=V = - \frac{i \l}{2} \sigma_3 +Y_0 =- \frac{i \l}{2}
\(\begin{array}{ccccc}
 1 &  0\\
 0 & -1 \end{array}\)  +
\(\begin{array}{ccccc}
  0& \overline{\psi} \\
 \psi & 0 \end{array}\)
$$
and 
$$
V_3(x,t) = \lt^2i \sigma_3 -\l Y_0 + |\psi|^2 i\sigma_3 -i \sigma_3 Y_0'.
$$
The corresponding   auxiliary linear problem     
$$
(\partial_x  -V )\f=0, \qquad\qquad\qquad    \f=\(\begin{array}{ccccc} f_1\\
                                                        f_2\end{array}\)
$$  
can be written in the form of an  eigenvalue problem for the Dirac operator 
$$
 \Dg \f= \[\(\begin{array}{ccccc} 1& 0 \\ 0 &  -1  \end{array}\) i\partial_x +
\(\begin{array}{ccccc}   0& -i \psib \\
 i \psi & 0 \end{array}\)\] \f= \lt \f.
$$

Another important feature of     soliton  systems is the Hamiltonian formulation. Here we assume that the potential $\psi(x,t)$ is 
$2l$--periodic: $\psi(x+2l,t)=\psi(x,t)$. 
For instance,  the NLS flow can be written as  
$$
\psi^{\bullet}= \{\psi, H_3\},
$$
with  Hamiltonian
$H_3=\frac{1}{ 2} \int_{-l}^{l} |\psi'|^2 +|\psi|^4 \, dx=${\it energy} and  bracket
$$
\{A,B\}=2i \int_{-l}^{l} \frac{\partial A}{ \partial \psib(x)}\frac{\partial B}{ \partial
\psi(x)}- \frac{\partial A}{ \partial \psi(x)}\frac{\partial B}{ \partial \psib(x)}\, dx.
$$The bracket is nondegenerate. 
The corresponding  symplectic form (up to a scalar) is: 
$$
\omega_0= 2i< \d \psib \wedge  \d \psi >, \qquad\qquad\qquad <\bullet>\;\;= \frac{1}{ 2l} \int_{-l}^{l} dx.
$$

{\it A priori} it  is not clear why the dynamical system, which arises as a compatibility
condition    has a    Hamiltonian formulation.   To put it 
differently, is it possible  to obtain   Hamiltonian formalism    from the spectral problem? 

Here we would like to make some historical remarks.  Originally, the Hamiltonian formulation of basic integrable models 
was found as an experimental fact. For the KdV  equation  
the computation of symplectic structure in terms of the scattering data was performed by Faddeev and 
Zakharov, \cite{FZ}. It  involved some
nontrivial identities for the products  of solutions. Later Kulish and Reiman,  \cite{KR},  noted that
all higher  symplectic structures also can be written in terms of the scattering data. Again, 
they used the scheme of \cite{FZ} and explicit calculations.  Finally,  we note that Zakharov and Manakov,  
\cite{ZM}, for the NLS equation  adopted a different approach.  Instead of the symplectic structure they worked with the corresponding    
Poisson bracket.  Again, using explicit formulas for the product of solutions they   computed the  
Poisson bracket between the coefficients of the scattering matrix. An  appearance  of  explicit formulas that are the moving force of all 
these computations seems to be quite mysterious. This  was already   discussed in  the literature  
\cite{FT},  and described as a  ''computational miracle".

The standard assumption needed to carry out spectral analysis is that the potential either is periodic 
or has rapid decay at infinity.  We refer to the latter case as scattering. 
Recently, in connection with the Seiberg-Witten theory \cite{SW1,SW2}, Krichever and Phong, \cite{KP}    
developed a  new approach  for the construction of  symplectic  formalism. The latest exposition of their results can be found in \cite{DKP}.   
The main idea of the  Krichever--Phong approach is to introduce in a universal 
way  the two-form on the   space 
of auxiliary linear operators. This form is  written in terms of the operator itself and its
eigenfunctions. The goal of this paper is to review  the Krichever-Phong  approach in the case of 1D periodic NLS  
and to extend  it  to the scattering case. Within the unified approach,  we  reduce the number  of formulas and eliminate unnecessary explicit computations. For instance computation of the symplectic form in terms of the spectral data (both in the periodic and the scattering case) becomes  
an application of the Cauchy residue theorem. 

\subsection{  The periodic case.}  We assume that the potential is periodic with the period $2l: 
\; \psi(x+2l,t)=\psi(x,t)$. The Krichever--Phong formula,  in the  NLS context,  takes the form   
$$
\omega_0=\sum\limits_{P_{\pm}} \R  <\e^* J\; \delta V \wedge \delta \e> d \lambda.
$$
This formula defines a closed 2--form $\omega_0$ on the space of operators
$\partial_x - V(x,\l)$ with $2l$ periodic potential.  The set--up for this formula is broadly as follows \footnote{ We refer to Section 2  for detailed discussion.}.

The eigenvalue problem $$[J\partial_x -JV(x,\l)]\e(x,\l)=0,\qquad\qquad\qquad J=i\sigma_2= \(\begin{array}{ccccc}
  0& 1 \\
 -1 & 0 \end{array}\)$$ 
has  special solutions, so--called Floquet solutions determined by  the property $\e(x+2l,\l)= w(\l) \e(x,\l)$. 
The complex constant $w(\l)$ is  called a Floquet multiplier.   For each value of the spectral parameter $\l$  
there are two linearly independent  Floquet solutions and  two  distinct Floquet multipliers. 
These solutions  and correspondingly multipliers become single--valued functions of a  point on the  two--sheeted covering of the  plane of 
spectral parameter $\l$.  The simple points of the periodic/antiperiodic spectrum of the eigenvalue problem constitute 
branching points of the cover. We assume that there is a finite number of simple points (so--called, finite gap potential). 
\begin{figure}[ht]
\includegraphics[width=0.60\textwidth]{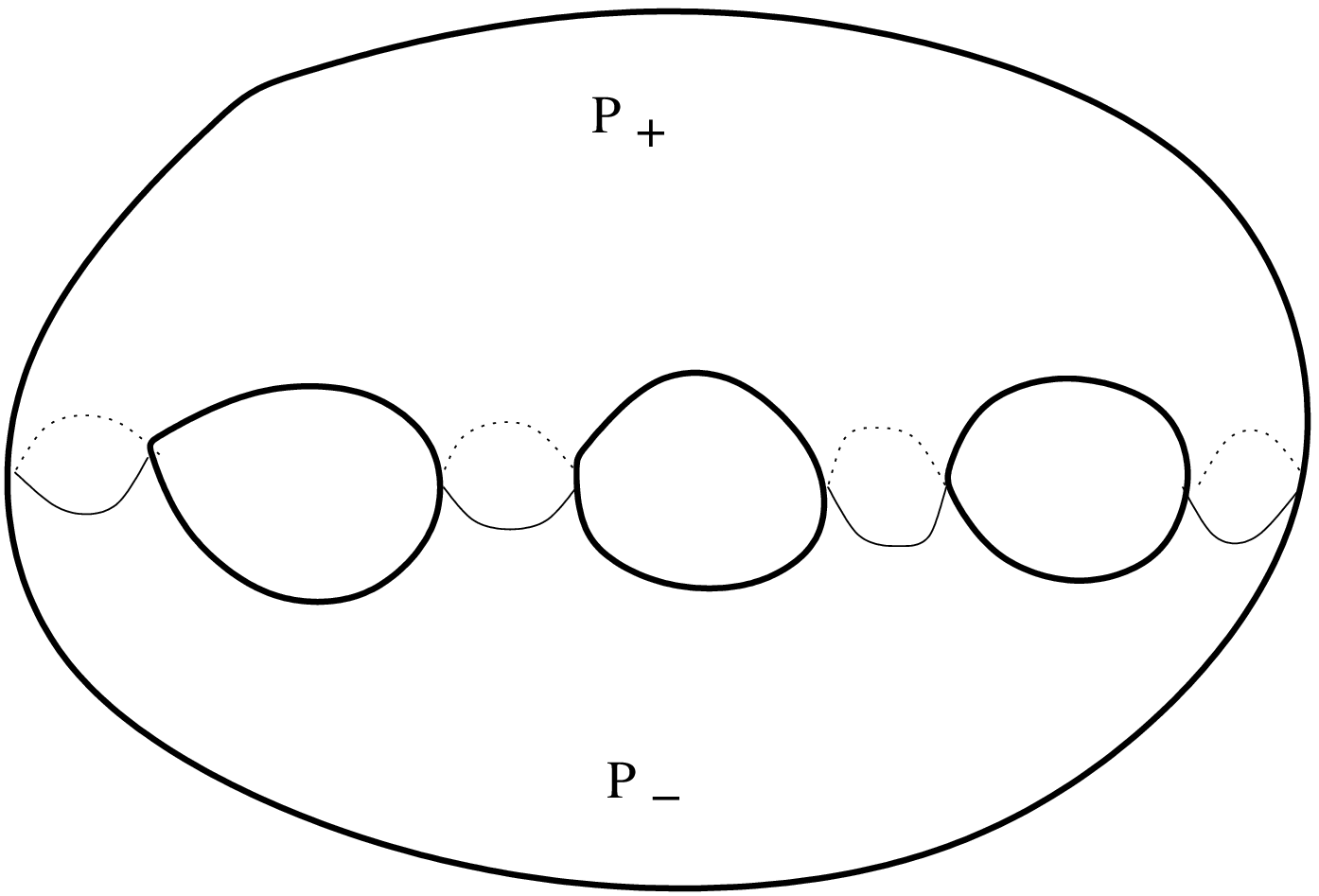}
\caption{Smooth Riemann surface $\Gamma$.}
\end{figure}

This  two sheeted covering constitutes   a hyperelliptic Riemann surface  $\Gamma$ with two infinities $P_{+}$ and $P_-$ (Fig. 1). Each point $Q=(\l,\pm)$ of  $\Gamma$ is specified by the value of  spectral parameter $\l$ and the sheet ''+'' or ''-'' which determines the Floquet multiplier $w(Q)$  corresponding to this $\l$. At every point of the curve we also have a Floquet solution $\e(x,Q)$ which  becomes a function of the point $Q$ and  satisfies the  
identity $\e(x+2l,Q)= w(Q) \e(x,Q)$. The Floquet solution  $\e(x,Q)$  has an exponential singularity at infinities and plays the role of   
so--called  Baker--Akhiezer function for the curve $\Gamma$. 

At every point of  the  curve $\Gamma$ we can define another solution $\e^*(x,Q)$. This is the Floquet solution 
which is brought  from  a  point on the different sheet but with the same  value of the spectral 
parameter $\l$. It is transposed and suitably normalized.   The  operator $J\partial_x -JV(x,\l)$ acts on the solution $\e^*(x,Q)$ as an adjoint {\it i.e.}  on the right:
$$
\e^*(x,Q)\[J\partial_x-J V\]=0. 
$$

It is   assumed    that the phase space consists of  smooth
$2l$--periodic functions $\psi(x)$ or equivalently operators $\partial_x - V(x,\l)$ with $2l$--periodic potential.
The  NLS flow   acts  on this space as it acts  on the space of functions $\psi(x)$.
All notions of differential geometry with obvious  conventions can be applied to this space of operators.   
On the space of potentials we have a  variation $\delta \psi(x)$. 
Thus for a fixed value of the spectral parameter  $\l$ we have well defined variation $ \delta V(x,\l)$.    The variations  $ \delta \e(x,Q), 
\delta \e^*(x,Q)$  are  defined correctly when $\l=\l(Q)$ is fixed. Therefore,  at each point $Q$ of the surface $\Gamma$ we have well defined meromorphic in  $Q$ the two--form 
$$
<\e^* J\delta V \wedge \delta \e> d \lambda.
$$
It takes values in the space of  skew--symmetric two--forms on  the space of operators 
$\partial_x - V$.  The result of Krichever and Phong states that the sum of residues of this form at infinities $P_{\pm}$ is nothing  but the symplectic 
form $\omega_0$.

The formula has a lot of good properties. First,  it  produces    all higher symplectic structures  by introducing the weight $\l^n$ under the residue
$$
\omega_n=\sum\limits_{P_{\pm}} \R \; \l^n\,<\e^* J\; \delta V \wedge \delta \e> d\,\lambda,\qquad\qquad\qquad\qquad n=1,2,\hdots.
$$
Second,  it easily  leads to the Darboux coordinates, or in physics terminology the separation variables, see  Sklyanin \cite{S}.
These are local coordinates where the symplectic form  $\omega_0$ takes the simple canonical form   
$$
\omega_0=\frac{2}{i}\sum\limits_{k} \d p(\gamma_k) \wedge \d \l (\gamma_k).
$$
This merits special explanation. 
It is well-known since the work of Flashka--McLaughlin, \cite{FM},   
that the poles $\gamma_k$ of  Floquet solutions lead to the Darboux coordinates for  symplectic 
forms \footnote{See also  Novikov--Veselov,  \cite{NV},
 for general discussion.}.   
Recently,   a lot of work was performed, \cite{SNK}, to construct such variables for the Ruijsenaars--Sneider 
and the Moser--Calogero systems.  This required formidable technical machinery and extensive 
computations.   At the same time,  as it was demonstrated by Krichever, \cite{K2},   the formula leads 
to the same result only  by applying the Cauchy residue theorem. 

\subsection{ The Scattering case.} The main goal of the present paper is to show that suitably interpreted
the new  approach  can be adopted for soliton systems with rapidly decaying initial data
on the entire line.  This is so--called the scattering case\footnote{ We refer to Section 3 for detailed definitions.}.

For such  potentials  one can define so--called Jost solutions $J_{\pm}(x,\l)$.  
These are matrix solutions of the auxiliary linear problem $J_{\pm}'= V J_{\pm}$ with the asymptotics 
$J_{\pm}'(x,\l)= \exp{(-{i\lt x}  \sigma_3)} + o(1)$, as $x\rightarrow \pm \infty$.   
Their columns  $J_{\pm}=\[\jo^{(1)}_{\pm},\jo^{(2)}_{\pm}\]$ are analytic in the corresponding 
upper/lower half--plane. 

\begin{figure}[htb]
\includegraphics[width=0.60\textwidth]{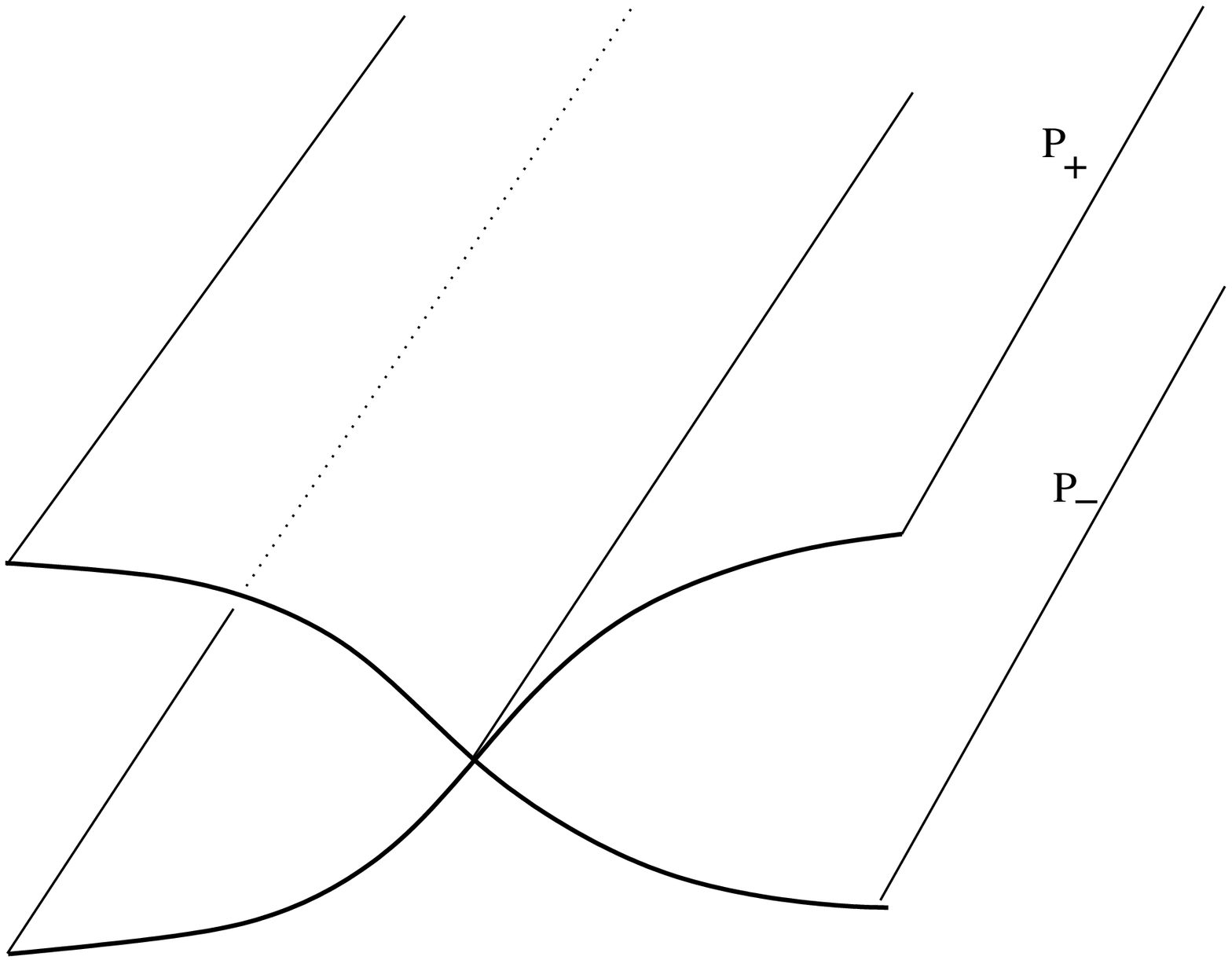}
\caption{ Singular Riemann surface $\Gin$.}
\end{figure}

Our construction of the associated Riemann surface $\Gin$ is a geometrical interpretation of what
is called the Riemann-Hilbert approach to the scattering problem, see \cite{FT}.
A singular curve $\Gin$ is obtained by taking two copies of the complex plane and gluing  them to each other along the real line (Fig. 2). 
The curve $\Gin$ has two infinities $P_+$ and $P_-$ and continuum set of singular points
above the real line.  The standard Jost solutions are lifted on
$\Gin$ and become the single valued function of a point on  the curve.  
Different branches of BA function are connected along the real line by  the
scattering matrix $S$:
$$
S(\l) = \frac{1}{ a} \[\begin{array}{ccccc} \;\; {1} & {\overline{b}}\\
                        -{b}& {1}
    \end{array}\].
$$
The Jost solution
has exponential singularity at infinities and plays the role of the  Baker--Akhiezer function for
the curve $\Gin$. This  construction is explained in detail in Section 3.

The formula of Theorem 3.4 looks similar to the periodic case
$$
\omega_0=
\tr\;  \R\;  \frac{1}{ 2} \[ <H_+^* J\delta V \wedge \delta H_+> +
  <H_-^* J\delta V \wedge \delta H_->\] d \lambda. 
$$ 
The only difference now is that we  work with the matrix solutions 
$$
H_+(\l)=\[ \jo_-^{(1)}(\l),\jo_+^{(2)}(\l)\] \quad {\rm and}\quad\quad H_-(\l)=\[ \jo_
+^{(1)}(\l),\jo_-^{(2)} (\l) \]
$$ 
and $H_+^+(\l)=\sigma_1 H_+^T$,   $H_-^+(\l)=\sigma_1 H_-^T$, with  
$$
\sigma_1=\left(\begin{array}{ccccc}
 0& 1\\
1 & 0
\end{array}  \right).
$$
The averaging now corresponds to the integration on the entire line
$$
<\bullet>\;\;=  \int_{-\infty}^{+\infty} dx. 
$$
The residue can be computed explicitly $\omega_0=2i<\d \psib\wedge \d \psi>$.
Theorem 3.6 states that the symplectic structure can be put in the Darboux form
$$
\omega_0= \frac{1}{ \pi i}  \int\limits_{-\infty}^{+\infty}
\frac{\d \bar b(\l)\wedge \d b(\l)}{ |a(\l)|^2}\, d\l,
$$
where $a$ and $b$ are  coefficients of the scattering matrix $S$.  
Again identically to the periodic case this result is obtained  by applying the Cauchy residue theorem. 
Only now  the sum of the residues in the affine part of the curve is  replaced by its'   continuous 
analog. This is the  integral which stays in the  right hand side of the formula.

The  unified approach to construction of  symplectic forms  produces an  interesting  problem.  As we   see,    
the  symplectic form constructed in the periodic case has two systems of Darboux coordinates. 
One system is associated with poles of the Floquet solution. It is   
the divisor--quasimomentum  Darboux coordinates. Another system of Darboux coordinates is the  
action-angle variables. At the same time  in the scattering case we know only one system of 
Darboux coordinates. These  are  action-angle variables. What is the correct analog of the 
divisor--quasimomentum in the scattering case? This is a subject of future publication, \cite{V2}.

We conclude this introduction by expressing thanks to  A. Its, H. McKean and I. Krichever for
stimulating discussions. We are also greatful to anonymous referee for remarks that helped to improve the presentation.

\section{ The Periodic Case.}
\subsection{The Direct Spectral Problem}

We provide here  information  needed in the next section for construction of symplectic forms.
We refer  
to classical  books \cite{MO,NMP} for standard facts of  spectral theory and   algebraic--geometrical 
approach to solitons. 

The  NLS equation
\bay\label{nls}
i \psi^{\bullet}= -\psi'' + 2 |\psi |^2 \psi, 
\ey
where $\psi(x,t)$ is  a  smooth complex function  $2l$--periodic in $x$,
is  a Hamiltonian system
$$
\psi^{\bullet}= \{\psi, H\},
$$
with the Hamiltonian
$H=\frac{1}{ 2} \int_{-l}^{l} |\psi'|^2 +|\psi|^4 \, dx=${\it energy} and the bracket
\bey
\{A,B\}=2i \int_{-l}^{l} \frac{\partial A}{ \partial \psib(x)}\frac{\partial B}{ \partial
\psi(x)}- \frac{\partial A}{ \partial \psi(x)}\frac{\partial B}{ \partial \psib(x)}\, dx.
\eey
The NLS Hamiltonian $H=H_3$ is one in the infinite series of conserved integrals of motion.
\bey
H_1&=&\frac{1}{ 2} \int_{-l}^{l} |\psi|^2 dx,\\
H_2&=&\frac{1}{ 2i} \int_{-l}^{l} \psi \bar{\psi}'  dx,\\
H_3&=&\frac{1}{ 2} \int_{-l}^{l} |\psi'|^2 +|\psi|^4 \, dx,\quad etc.
\eey
These Hamiltonians produce an infinite hierarchy of  flows $e^{tX_m},\; m=1,2,\ldots$.

The first in the  hierarchy is the  phase flow $e^{tX_1}$  generated by the vector field
$$
X_1:\qquad\qquad \psi^{\bullet}=\{\psi, H_1\}=  -i \psi.
$$
The phase flow is a compatibility condition for
\bay\label{com}
[\partial_t- V_1,  \partial_x-V_2]=0, 
\ey
with\footnote{
Here and below  $\sigma$  denotes the {\it Pauli matrices}
$$
\sigma_1=\left(\begin{array}{ccccc}
 0& 1\\
1 & 0
\end{array}  \right),  \nonumber \quad
\sigma_2=\left(\begin{array}{ccccc}
0 & -i\\
i &  0
\end{array} \right), \nonumber \quad
\sigma_3=\left(\begin{array}{ccccc}
1 & 0\\
0 & -1
\end{array} \right). \nonumber
$$
}
$V_1=\frac{i}{ 2} \sigma_3$ and 
$$
V_2 = - \frac{i \l }{ 2} \sigma_3 +Y_0 =
\(\begin{array}{ccc}
 - \frac{i\l}{ 2} &  0\\
 0 & \frac{i\l}{ 2} \end{array}\)  +
\(\begin{array}{ccccc}
  0& \overline{\psi} \\
 \psi & 0 \end{array}\).
$$
We often omit the subscript $V=V_2$.
The second, translation flow $e^{tX_2}$ generated by 
$$
X_2:\qquad\qquad \psi^{\bullet}=\{\psi, H_2\}=   \psi'
$$
is equivalent to \ref{com}  with $V_1$ replaced by $V_2$.
Finally, the third, original NLS flow \ref{nls} is a compatibility condition for \ref{com} 
with $V_1$ replaced by 
$$
V_3 = \frac{\l^2}{ 2}i \sigma_3 -\l Y_0 + |\psi|^2 i\sigma_3 -i \sigma_3 Y_0'.
$$

All flows of infinite hierarhy  $e^{tX_m},\; m=1,2,\ldots$  commute  with each other  
$$
[\partial_{\tau_m} - V_m,  \partial_{\tau_n} -V_n]=0.
$$
The first times $\tau_1,\tau_2$ and $\tau_3$ correspond to the first three flows.

We introduce a  $2\times 2$  transition  matrix $M(x,y,\l),\; x \geq y$; that satisfies  
$$
M'(x,y,\l)=V(x,\l) M(x,y,\l),\qquad\qquad\qquad M(y,y,\l)=I.
$$
The solution is given by the formula
$$
M(x,y,\l)=\exp\int_{y}^{x} V(\xi,\l) d\xi.
$$
The   matrix $M(x,y,\l)$ is unimodular  because $V$ is traceless.  

The symmetry 
$$
\sigma_1 \overline{V}(x,\l)\sigma_1= V(x,\bar{\l})
$$
produces  the same relation for the  transition matrix 
\bay\label{real}
\sigma_1 \overline{M}(x,y,\l)\sigma_1= M(x,y,\bar{\l}).
\ey
Another symmetry
$$
V^T(x,\l) J=-J V(x,\l),
$$
where $J=i\sigma_2$, implies 
\beq\label{nsy}
M^T(x,y,\l)^{-1}J=J M(x,y,\l).
\eeq

 The quantity  $\Delta(\l)=\frac{1}{ 2} \text{trace}  M(l,-l,\l)$  is called a discriminant.
The formula \ref{real} implies  $\overline{\Delta}(\l)=\Delta(\bar{\l})$ and $\Delta(\l)$ is real for real $\l$. 
The eigenvalues of the monodromy matrix  have a name of  Floquet multipliers and 
they are  roots of the quadratic equation 
\beq\label{qe}
 w^2 -2\Delta w +1=0.
 \eeq
The Floquet multipliers are  given by the formula $w=\Delta\pm \sqrt{\Delta^2-1}$. The values of $\l: w(\l)=\pm 1$ constitute the points of the periodic/antiperiodic 
spectrum.  The corresponding    auxiliary linear problem 
$$
(\partial_x  -V )\f=0, \qquad\qquad\qquad    \f^T=(f_1,f_2);
$$
can be written in the form of an  eigenvalue problem for the self--adjoint Dirac operator
$$
{\Dg} \f= \[\(\begin{array}{ccccc} 1& 0 \\ 0 &  -1  \end{array}\) i\partial_x +
\(\begin{array}{ccccc}
 0& -i \overline{\psi} \\
 i \psi & 0 \end{array}\)\] \f= \lt \f.
$$
The self--adjointness implies that points of the spectra are real.

\noi
{\it Example.} Let $\psi \equiv 0$. The corresponding monodromy matrix  can be easily computed
$M(x,y,\l)=e^{-{i\lt}\sigma_3 (x-y)}$. 
We have $\Delta(\l)=\cos \l l$ and double
eigenvalues at the points $\l_n^{\pm}=\frac{\pi n}{  l}.$
If n is even/odd, then the corresponding $\l_n^{\pm}$ belongs to the  periodic/
anti-periodic spectrum.

For a generic potential the double points $\l_n^{\pm}$ of the periodic/anti-periodic 
spectrum split, but they always stay real.  
The size of the spectral gap is determined, roughly speaking, by
the corresponding Fourier coefficients of the potential. In our considerations we assume that 
there is a finite number of $g+1$  open  gaps in the spectrum 
$$
\hdots<\l_{n-1}^{-}=\l_{n-1}^{+}<\l_{n}^{-}<\l_{n}^{+}<\hdots< \l_{n+g}^{-}<\l_{n+g}^+< \l_{n+g+1}^{-}=\l_{n+g+1}^{+}<\hdots
$$
These are so--called finite gap potentials which are dense among all potentials. 

The Floquet multipliers  become single--valued on
the Riemann surface:
$$
\Gamma=\{Q=(\l,w)\in { \C}^2:\quad   R(\l,w)=\det\[M(l,-l,\l)-wI\]=0\}.
$$
The Riemann surface consists of two sheets covering the plane of the spectral parameter $\l$. 

\noi
{\it Example.} Let $\psi \equiv 0$.  
We have $\Delta(\l)=\cos \l l$ and quadratic equation \ref{qe} has the  solutions $w(\l)=e^{\pm il \l}$. 
The Riemann surface $\G=\G_+ + \G_-$ is reducible and consists of two copies of the complex plane $\C$ that intersect each other at the points of the double spectrum   $\l_n^{\pm}$.  Each part  $\G_+$ or $\G_-$ contains the corresponding infinity $P_+$ or $P_-$. 
The Floquet multipliers are  single valued on $\G$: 
\bey
w(Q)&=&e^{+i\l l},\qquad\qquad\qquad\qquad Q\in \G_+;\\
w(Q)&=&e^{-i\l l},\qquad\qquad\qquad\qquad Q\in \G_-.
\eey

For a finite--gap potential the Riemann surface $\G$ is irreducible. 
There are three types of important points on $\Gamma$. These are the singular points, the points above $\l=\infty$ and the branch points which we 
discuss now in detail. 
\begin{itemize}
\item The singular points are determined by the condition
$$
\partial_{\l}R(\l,w)=\partial_{w}R(\l,w)=0.
$$
These are the points $(\l^{\pm},\pm 1)$ of the double spectrum. At these points two sheets of the curve intersect. 
\item There are two nonsingular points $P_+$ and $P_-$ above $\l=\infty$. At these points\footnote{The notation $Q\in (P)$ means that the point $Q$ is in the vicinity  of the point $P$.}  
\bay
w(Q)&=&e^{+i\l l}\(1+O\({1/\l}\)\),\qquad\qquad Q\in (P_+);\label{asf}\\
w(Q)&=&e^{-i\l l}\(1+O\({1/\l}\)\),\qquad\qquad Q\in (P_-).\label{assf}
\ey
\item The branch points are specified by the condition
$$
\partial_{w}R(\l,w)=0.
$$
They are different from the singular points and correspond to the simple  periodic/antiperiodic spectrum. We denote these points by  
$s_k^{\pm}=(\l^{\pm}_k,(-1)^k),$ $\;k=n,\hdots,n+g$. There are $2(g+1)$ of them, each 
has a  ramification index 2. 
\end{itemize}
\noi
The  desingularized curve $\G$ is biholomorphicaly equivalent to a  hyperelliptic curve with branch points at the points of the simple spectrum. We also 
 denote the hyperellitic curve by $\G$. 
The Riemann-Hurwitz formula for the genus of $\G$ implies 
$$
\text{genus}=\frac{R}{2}-n+1,
$$
where $R$ is a total ramification index  and $n$ is  the number of sheets. Each branch point has a ramification index $1$ and therefore $R=2(g+1)$ and 
$n=2$. Therefore,  the genus of $\G$ is $g$, one off the number of open gaps in the spectrum.

 Let  $\epm$ be   a 
holomorphic involution on the curve $\Gamma$ permuting sheets 
$$\epm: \; (\l,w)\longrightarrow (\l,1/w).$$ 
The fixed points of $\epm$ are the branch points of $\G$. The involution $\epm$ permutes infinities $\epm: P_-\longrightarrow P_+$.
Let us also  define on $\Gamma$  an antiholomorphic involution  
$$\epsilon_a: (\l, w ) \longrightarrow (\bar{\l}, \bar{w}).$$ The involution $\ea$ also permutes infinities and commutes with $\epm$.
Points of the curve above  gaps $\[\l_n^-, \l_n^+\]$ where $|\Delta(\l)| \geq 1$ form  $g+1$  fixed ``real'' ovals of $\epsilon_a$.   
We call them $a$--periods. 

The  quasimomentum $p(Q)$ is a multivalued function on the curve $\Gamma$. It  is introduced 
by the formula $w(Q)=e^{i p(Q)2l}$. Evidently,  it is defined up to $\frac{\pi n}{ l}$, where $n$ is an integer. 
The asymptotic expansion for $p(Q)$ at infinities can be easily computed
$$
\pm p(\l)=  \lt - p_{0}^{\pm} -  \frac{p_{1}}{ \l} - \frac{p_{2}}{ \l^2}\ldots,
\quad \quad \quad \quad Q\in(P_{\pm}),\;\, \l=\l(Q);
$$
where $p_{0}^{\pm}  = \frac{\pi k_{\pm}}{ l}, \; k_{\pm}$ is an integer and
$$
p_{1}  =   \frac{1}{ l} H_1, \quad
p_{2}  =   \frac{1}{ l}  H_2, \quad
p_{3}  =   \frac{1}{ l} H_3, \quad etc.  
$$
Moreover,  the function $w(Q)+ w(\epsilon_\pm Q)$ does not depend on the sheet and  is equal to $2\Delta(\l)$.  
Thus $\Delta(\l(Q))=\cosh i p(Q) 2l$ and the formula 
$$
dp= \pm \frac{1}{ i 2l} d \cosh^{-1} \Delta(\l) = \pm \frac{1}{ i 2l}\frac{\Delta^{\bullet}(\l)\,
d\l}{  \sqrt{\Delta^2-1}}
$$
implies that  differential $dp$ is of the second kind with double poles at the infinities:
$\pm dp=d\(\lt +O(1)\)$. The same formula implies that the differential $dp$ is pure complex on the real ovals. At the same time, the condition  $w(s_k^-)=w(s_k^+)$ requires   the increment $p(s_k^+)-p(s_k^-)$ to be real.   
 Therefore,  $dp$   has  zero $a$-periods 
$$
\int_{a_k} dp=0.
$$
Since the Floquet multiplies are single--valued on $\G$ for the $b$--periods we have 
\beq\label{pc}
\int_{b_k} dp=\frac{\pi n_{b_k}}{ l},\;\qquad\qquad\qquad n_{b_k} \in \Z,\;\quad k=1,\hdots,g.
\eeq
These are  so--called the {\it periodicity conditions}, \cite{NMP}.

The Floquet solution is the  vector--function
$$
\e(x,Q)= \[\begin{array}{ccccc}
e^1(x,Q)\\
e^2(x,Q)\end{array} \]
$$
which is a solution of the auxiliary spectral problem $\e'= V \e$ with the property
\beq\label{mp}
\e(x+2l, Q) = M(l,-l,\l) \e(x,Q)= w(Q) \e(x, Q)
\eeq
 and normalized by the condition 
\bay\label{norm}
e^1(-l,Q) +e^2(-l,Q)=1.
\ey

\noi
{\it Remark.}  If $\f(x,\l)$ is a solution of the auxiliary problem 
$$
(\partial_x  -V(x,\l) )\f=0
$$
corresponding to $\l$, then $\hat \f=\sigma_1 \bar{\f}$ is a solution of $(\partial_x  -V(x,\overline{\l}) )\hat\f=0$ corresponding to $\overline{\l}$. 

\noi
{\it Example.} Let $\psi=0$.  The Floquet solution is
given by the formula
$$
\e(x,Q) = e^{+i \lt(x+l)} \e_0 = e^{+i \lt(x+l)} 
\[\begin{array}{ccccc}  0\\ 1\end{array}\], \qquad\qquad Q\in \G_+,
$$
$$ 
\e(x,Q) = e^{-i \lt(x+l)} \hat \e_0 = e^{-i \lt(x+l)} \[\begin{array}{ccccc}
1\\
0\end{array} \],   \qquad\qquad Q\in \G_-.
$$
It has  no poles in the affine part of the curve.

For a general finite gap potential the situation is more complicated. 
\begin{lem}
\label{lfloc}The Floquet solution satisfies the identity
$$
e(x,\ea Q)=\sigma_1 \overline{\e(x,Q)}.
$$
The Floquet solution $\e(x,Q)$ has poles common for both components at the points $$\gamma_1,\gamma_2,\hdots,\gamma_{g+1}.$$ 
Projections of poles $\mu_k= \l(\gamma_k)$ are  real. Each $\gamma_k$ lies on the real oval above the corresponding open
gap $[\l_k^-,\l^+_k]$. 
Each component  $e^i(x,Q)$ has  $g+1$   zeros $$\sigma^i_1(x),\; \sigma^i_2(x),\hdots,\sigma^i_{g+1}(x);\qquad\qquad\qquad\; i=1,2.$$ These zeros depend on the parameter $x$.
In the vicinity of infinities  the function $\e(x,Q)$ has the asymptotics
$$
\e(x,Q) =  e^{\pm i \lt(x+l)} \[ \e_0 / \hat \e_0 + o(1)\], \qquad \qquad
Q\in (P_{\pm}).
$$
\end{lem}

Before proceeding to the proof of the Lemma we note that the differential equation for the monodromy matrix    
$$
M'(x,y,\l)=\[- \frac{i \l }{ 2} \sigma_3 +Y_0\] M(x,y,\l),\qquad\qquad\qquad M(y,y,\l)=I,
$$
multiplied (gauged) on the left and right by the matrices
$$
C=\(\begin{array}{ccccc} 1& 1 \\ i &  -i  \end{array}\)\qquad\text{and}\qquad 
C^{-1}=\frac{1}{2}\(\begin{array}{ccccc} 1& -i \\ 1 &  i  \end{array}\)
$$
transforms into
$$
M^{\RB}(x,y,\l)=\[- \frac{i \l }{ 2} \sigma_2 +Y_0^{\RB}\] M^{\RB}(x,y,\l),\qquad\qquad\qquad M^{\RB}(y,y,\l)=I;
$$
with 
$$
Y_0^{\RB}=\(\begin{array}{ccccc} q& p \\ p &  -q  \end{array}\),\qquad\qquad\qquad \psi=q+ip.
$$
This is a real version of the eigenvalue problem which is more convenient in some situations, see \cite{MV}. The Floquet solution 
 $\e^{\RB}(x,Q)$ corresponding to the real version of the eigenvalue problem is related to $\e(x,Q)$ by the formula 
$$\e^{\RB}(x,Q)=C \e(x,Q).
$$
Therefore the result of the Lemma for $\e(x,Q)$ follows from the corresponding result for $\e^{\RB}(x,Q)$ given in \cite{MV}. We prefer to give 
 a direct proof, though the gauge transformation is behind all arguments.

\noi
{\it Proof.} The proof is based on the explicit formula for the Floquet solution. Let  
$$
M(x,-l,\l)= \[  \begin{array}{ccccc}
                 m_{11} &  m_{12}\\
                m_{21} &  m_{22} \end{array}\] (x,-l,\l);
$$
and $M_{11}, M_{12},$ {\it etc.} be  the elements of the matrix $M(l,-l,\l)$.
The Floquet solution $\e(x,Q)$ is given by the formula  
\bay\label{floq}
\e(x,Q)=A(Q) \[\begin{array}{cccc}
m_{11}\\
m_{21} \end{array} \](x,-l,\l) +   (1-A(Q)) \[\begin{array}{cccccc}
m_{12}\\
m_{22} \end{array} \](x,-l,\l),
\ey
where $\l= \l(Q)$  and  the coefficient $A(Q)$ is 
\bay\label{coef}
A(Q)=\frac{M_{12}}{ M_{12}-M_{11} +w(Q)} \quad \quad \text{or}  \quad \quad
A(Q)=\frac{w(Q)- M_{22}}{ M_{21}- M_{22} +w(Q)}.
\ey

To prove the formula note that the Floquet solution is a linear combination of  columns of the  monodromy matrix $M(x,-l,\l)$:
$$
\e(x,Q)=A(Q) \[\begin{array}{cccc}
m_{11}\\
m_{21} \end{array} \](x,-l,\l) +   A'(Q) \[\begin{array}{cccccc}
m_{12}\\
m_{22} \end{array} \](x,-l,\l),   \quad \quad \l= \l(Q).
$$
The normalization condition \ref{norm} implies $A'(Q)=1-A(Q)$. At the same time the Floquet solution is an eigenvector of the monodromy matrix
$$
M(l,-l,\l)  \[\begin{array}{cccc}
A(Q)\\
1-A(Q) \end{array} \]= w(Q) \[\begin{array}{cccc}
A(Q)\\
1-A(Q) \end{array} \].
$$
This  leads to two equations 
$$
M_{11}A(Q)  + M_{12} (1-A(Q))=w(Q) A(Q),
$$
or
$$
M_{21}A(Q)+ M_{22}(1-A(Q))=w(Q) (1-A(Q)).
$$
Each equation implies the corresponding formula for $A(Q)$. 

The formulas \ref{real} and \ref{coef} imply
$$
1-A(\ea Q)=\overline{A(Q)}.
$$
This and  \ref{real},  \ref{floq}  imply the stated identity for the Floquet solution. 

The relation $M^{\RB}=CMC^{-1}$ implies 
\bey
M^{\RB}&=& \[\begin{array}{ccccc}  
                 M^{\RB}_{11} &  M^{\RB}_{12}\\
                 M^{\RB}_{21}&  M^{\RB}_{22} \end{array}\]\\
                 &=&\frac{1}{2} \[ \begin{array}{ccccc}  
                 M_{11}+M_{12}+M_{21}+M_{22}&  i(M_{12}+M_{22}-M_{11}-M_{21})\\
                 i(M_{11}+M_{12}-M_{21}-M_{22})&  M_{11}+M_{22}-M_{12}-M_{21} \end{array}\]. 
\eey
Due to \ref{real} $M^{\RB}(\l)$ is real for real $\l$. Consider the function $M^{\RB}_{12}(\l)$ and    look at the roots 
$\mu_n:\;M^{\RB}_{12}(\mu_n)=0$. For $\psi\equiv 0$ we have 
$M^{\RB}_{12}(\l)=-\sin\frac{\l l}{2}$ with roots at the points $\mu_n=\frac{2\pi n}{l},\; 
n\in \Z$.  When we add the potential the roots $\mu_n$ move  but stay real. They are caught by open gaps or match double 
periodic/antiperiodic spectrum. Indeed at $\mu_n$ the matrix $M^{\RB}$ is lower triangular and real entries $M^{\RB}_{11}$ and $M^{\RB}_{22}$ 
coincide with Floquet multipliers. Since $M^{\RB}_{11}M^{\RB}_{22}(\mu_n)=1$ we have $|\Delta(\mu_n)|=\frac{1}{2}|M^{\RB}_{11}+M^{\RB}_{22}|
\geq 1$.

The points of the divisor $\gamma_k\in \Gamma$ lie above the points $\mu_k$ on the sheet with $w(Q)=M^{\RB}_{22}(\mu_n)$. At these points
the denominator in \ref{coef}
$$
M_{12}-M_{11} +w(Q)\qquad\text{or}\qquad M_{21}- M_{22} +w(Q) 
$$
vanishes. Indeed from  $M^{\RB}_{12}(\mu_n)=0$ we have
$$ M_{12}-M_{11}(\mu_n)=M_{21}-M_{22}(\mu_n)$$
and $w(Q)=M^{\RB}_{22}(\mu_n)=M_{11}-M_{12}(\mu_n)=M_{22}-M_{21}(\mu_n)$. Moreover $M_{22}\neq M^{\RB}_{22}(\mu_n)$. These 
produce a pole of the Floquet solution when $\mu_n$ lies in the open gap. When $\mu_n$ is caught by the periodic/antiperiodic spectrum the matrix 
$
M(l,-l,\mu_n)=\pm I
$
and the zero of denominator is annihilated by the zero of numerator in \ref{coef}.

The asimptotics of the Floquet solution  follows from the formula \ref{floq} and
$$
M(x,y,\l)=e^{-{i\lt}\sigma_3 (x-y)}+ o(1),
\qquad\qquad \text{when}\qquad \l \rightarrow \infty. 
$$
 \qed

The Floquet solution $\e(x,Q)$ near  infinities can be expanded into the asymptotic series 
$$
\e(x,Q)=e^{+i \lt(x+l)}  \sum\limits_{s=0}^{\infty} \e_s(x) \l^{-s} =
e^{+i \lt(x+l)} \sum\limits_{s=0}^{\infty} \[\begin{array}{ccccccc}
b_s \\
d_s
\end{array} \] \l^{-s}, \quad   Q\in (P_{+}),
$$
$$
\e(x,Q)=e^{-i \lt(x+l)} \sum\limits_{s=0}^{\infty} \hat \e_s(x) \l^{-s}
= e^{-i \lt(x+l)} \sum\limits_{s=0}^{\infty} \[\begin{array}{ccccc}
\overline{d}_s \\
\overline{b}_s
\end{array} \] \l^{-s} ,  \quad Q\in (P_{-}),
$$
and  $b_0= 0,\quad d_0=1$. 
The coefficients $b_s, d_s$  can be computed from the  relation
\bay\label{rec}
-\[\begin{array}{ccccccc}
b_s'\\
d_s'
\end{array} \] + Y_{0}
\[\begin{array}{ccccccc}
b_s\\
d_s
\end{array} \]= \frac{i}{ 2} \(I +  \sigma_3 \)
\[\begin{array}{ccccccc}
b_{s+1}\\
d_{s+1}
\end{array} \], \quad\quad s=0,1,\ldots,
\ey
due to the diagonal form of the matrix 
$$
\frac{i}{ 2} \(I +  \sigma_3 \)= \[\begin{array}{ccccccc}  i& 0\\ 0 & 0 
\end{array} \].
$$
Indeed,   relation \ref{rec} leads to the identities
\bay\label{fir}
-b_s'  + \psib d_s = ib_{s+1},  
\ey
and
\bay\label{sec}
-d_s'  + \psi b_s = 0.   
\ey
These are supplemented  by the boundary condition
\bay\label{bo}
 b_s(x) + d_s(x)|_{x=-l}=0,\quad \quad \quad  s \geq 1. 
\ey

When $s=0$ using  $b_0= 0,\;\; d_0=1$   from \ref{fir}  we obtain 
$$
b_1=-i\overline{\psi}(x).
$$
The identities \ref{sec} and  \ref{bo}  imply
$$
d_1=i\overline{\psi}_0 - i \int\limits_{-l}^{x} |\psi|^2 dx',\qquad\qquad\qquad \psi_0=\psi(-l).
$$
Similar, we compute
\begin{eqnarray*}
b_2& =& \overline{\psi}' + \overline{\psi}\, \overline{\psi}_0 - \overline{\psi}
\int\limits_{-l}^x |\psi|^2 dx', \\
d_2&=& -\overline{\psi}_0' -  \overline{\psi}_0^2 + \int\limits_{-l}^{x}
\[ \psi \overline{\psi}' +|\psi|^2 \overline{\psi}_0-|\psi|^2 \int\limits_{-l}^{x'}
|\psi|^2 dx''\] dx'.
\end{eqnarray*}
These formulae will be used for   explicit computation of symplectic forms.

The Floquet solution $\e(x,Q)$ satisfies the identity
$$
\[J\partial_x- J V\] \e(x,Q) =0, \qquad\qquad\qquad J=i\sigma_2; 
$$
which is just another way to write the spectral problem.
Let us define the dual Floquet  solution $\e^+(x,Q)=\[e^{1+}(x,Q), e^{2+}(x,Q)\]$ at the point
$Q$ as
$$
\e^+(x,Q)=\e(x,\epsilon_{\pm }Q )^T.
$$
It can be verified by a direct computation that  the dual  Floquet solution 
$\e^+(x,Q)$ satisfies\footnote{The action of the differential operator 
$D=\sum\limits_{j=0}^{k} \omega_{j}
\partial^{j}$ on the row vector $f^{+}$ is defined as
$f^+ D=\sum\limits_{j=0}^{k} (-\partial)^{j}(f^{+} \omega_j).$}
$$
\e^+(x,Q)\[J\partial_x-J V\]=0.
$$
The fact that the Wronskian $\e^+(x,Q)J \e(x,Q)$ does not depend on $x$ can be verified by differentiation.   
Introducing the function  
$$
\F(Q)=\e^+(x,Q)J \e(x,Q)=<\e^+(x,Q)J\e(x,Q)>= \frac{1}{ 2l} \int_{-l}^{l}\e^+(x,Q)J\e(x,Q) dx,
$$
we define  another dual Floquet solution $\e^*(x,Q)$ by the formula
$$
\e^*(x,Q)=\frac{\e^+(x,Q)}{ \F(Q)}.
$$
Evidently,   $\e^*(x,Q) J \e(x,Q)=1$. The symmetry \ref{nsy}  produces an analog of monodromy property \ref{mp} for the 
function $\e^*$:
\beq\label{mps}
\e^*(x+2l, Q) = \e^*(x, Q) JM^{-1}(l,-l,\l)J^{-1} = w^{-1}(Q) \e^*(x, Q).
\eeq
\begin{lem} 
\label{lsfloq}
The function $\e^*(x,Q)$ has simple poles at the branch points
$s_k^{\pm}$. It has fixed zeros at $\gamma_1,\ldots, \gamma_{g+1}$.
The other zeros for each  component of the vector function $\e^*$ lie  on every real oval and depend on the parameter $x$. The function
$\e^*$ has the asymptotics at infinities
$$
\e^*(x,Q)=\pm e^{\mp \frac{i\l}{ 2}(x+l)} \[ {\hat \e}_0^T / \e_0^T +o(1)\],
\quad\quad Q\in(P_+/P_-).
$$
\end{lem}

\noindent
{\it Proof.} The function $\F(Q)$ is meromorphic with $2(g+1)$ poles on both sheets above points $\mu_n$ lying in open gaps and $2(g+1)$ 
zeros at the branching points $s_k^{\pm}$. At infinities it has the asymptotics 
$$
\F(Q)=\pm 1 +o(1),\qquad\qquad\qquad\qquad  Q\in (P_{\pm}).
$$
Now it is easy to prove properties of the function $\e^*(x,Q)$. It has poles at the branch points $s_k^{\pm}$ which arise from zeros 
of $\F(Q)$. It has zeros at $\gamma_1,\ldots, \gamma_{g+1}$, the poles of $\F(Q)$. Other poles of $\F(Q)$ are annihilated by the 
poles of $\e^+(x,Q)$. Other $g+1$ zeros of $\e^*(x,Q)$ which depend on $x$ are produced by the corresponding zeros of $\e^+(x,Q)$.

The asymptotics follows from the asymptotics for $\e(x,Q)$ and $\F(Q)$. 
\qed

Consider periodic variations of the matrix $V(x,\l): \tilde{V}= V+  \d V$. Then $\tilde{p}(Q)=p(Q)+  \d p(Q)+ \hdots$.
We need a standard formula connecting the variations   $ \d p(Q)$ and $\d V$.
\begin{lem} The following identity holds
$$
i\d p(Q)= < \e^*(x,Q)J \d V \e(x,Q)>.
$$
\end{lem}

\noindent
{\it Proof.}   Let $\tilde{\e}(x,Q)$ be  a Floquet solution corresponding  to the deformed potential $\tilde{V}$. 
From the definition 
$$
 \e^+(x,Q)\([J\partial_x -J \tilde{V}]\tilde{\e}(x,Q)\)=0
$$
and
$$
 \(\e^+(x,Q)[J\partial_x -J {V}]\)\tilde{\e}(x,Q)=0.
$$
Subtracting one identity from another, we have 
$$
\e^+(J\partial_x \tilde{\e})-(\e^+J\partial_x) \tilde{\e}=\e^+J \tilde{V}\tilde{\e} - \e^+J{V}\tilde{\e}= \e^+J\d {V}\tilde{\e}.
$$
Integrating both sides, we have 
$$
\int_{-l}^{+l}  \e^+J \tilde{\e}'+ \e^{+'}J \tilde{\e} \;dx=\e^+J \tilde{\e}|_{-l}^{+l}.  
$$
Using the identities
$$
\e^+(l,Q)=e^{-ip(Q)2l} \e^+(-l,Q)
$$ 
and 
$$
\tilde{\e}(l,Q)=e^{i\tilde{p}(Q)2l} \;\,\tilde{\e}^+(-l,Q)
$$
for the LHS, we have 
$$
\( e^{i\tilde{p}(Q)2l}e^{-ip(Q)2l}-1\) \e^+(x,Q)J \tilde{\e}(x,Q)= i\d p(Q)2l \F(Q)+ \text{lower order terms}.
$$
The RHS is equal to 
$$
2l < \e^+(x,Q)J \d V \e(x,Q)>+ \text{lower order terms}.
$$
Collecting leading terms, we obtain the stated identity.
\qed

Consider a real hyperelliptic spectral curve  $\Gamma$ of finite genus  corresponding to some periodic
potential $\psi$.
Let us introduce the Baker-Akhiezer  function $\e(\tau,x,t,Q)$ which depends  on  three parameters (times)
$\tau,\;x$ and $t$ and has the asymptotics at infinities
$$
\e(\tau,x,t,Q)=e^{\pm i\(-\frac{1}{ 2}\tau +{\lt}x-{\lt^2}t\)} \times\[ 
\e_0/ \hat \e_0 +o(1) \],\qquad \qquad Q\in (P_+/P_-). 
$$
The BA function  has poles at the points $\gamma$'s, located on the real ovals. These
properties define  the BA function uniquely. The BA function  can be written explicitly in 
terms of theta-functions of the curve $\Gamma$, \cite{K1}. 
The BA function  has Bloch  property in $x$-variable
$\e(\tau, l,t,Q)=w(Q) \e(\tau, -l,t,Q)$ and satisfies the identities
$$
\begin{array}{cccccc}
&\[J\partial_{\tau}-J V_1(\tau,x,t)\]\e(\tau,x,t,Q)=0,\\
&\[J\partial_{x}-J V_2(\tau,x,t)\]\e(\tau,x,t,Q)=0,\\
&\[J\partial_{t}-J V_3(\tau,x,t)\]\e(\tau,x,t,Q)=0.
\end{array}
$$
The   three  matrices  $V_1, V_2$ and $V_3$ are given at the beginning of this section.

Let us define the dual BA function $\e^{+}(\tau, x,t,Q)$ at the point $Q$ as
$$
\e^{+}(\tau, x,t,Q)\equiv \e(\tau, x,t,\epsilon_\pm Q)^{T}.
$$
The identity $w(Q)w(\epsilon_\pm Q)=1$ implies $\e^{+}(\tau,l,t,Q)=w(Q)^{-1} 
\e^{+}(\tau,-l,t,Q)$. 
The dual BA function $\e^{+}(\tau,x,t,Q)$ satisfies dual  identities
$\e^{+}(\tau,x,t,Q)\[J\partial_{\tau}-J V_1\]=0,$ {\it etc}.

\subsection{ Symplectic Structures. }

We assumed in  the previous section that the phase space consists of  smooth  
$2l$--periodic functions $\psi(x)$. Instead we can change the  language and think about the phase space 
as a space of operators $\partial_x - V_2$ with $2l$--periodic potential.  
The flows of the NLS hierarchy  act on this space as well they act on the space of functions $\psi$. 
All notions of differential geometry can be applied to this space of operators with 
obvious conventions.

\begin{lem}  The formula
$$
\omega_0=\sum\limits_{P_{\pm}} \R \; <\e^* J\delta V \wedge \delta \e> d \lambda,
$$
defines a closed  2--form $\omega_0$ on the space of operators
$\partial_x - V_2$ with periodic potential.
 The flows $e^{tX_m},\; m=1,2,\ldots$ on the space of operators defined by the
formula   
$$
[\partial_{\tau_m} - V_m,  \partial_x -V_2]=0,
$$
are Hamiltonian with the symplectic structure $\omega_0$ and the Hamiltonian
function $H_m$ (up to unessential constant factor).
\end{lem}

\noi
{\it Remark.}  The formula
$$
\omega_n=\sum\limits_{P_{\pm}} \R \;\l^n <\e^* J\delta V \wedge \delta \e> d \lambda,
\quad \quad \quad \quad n=0, 1 \ldots,
$$
defines a closed 2--form $\omega_n$ on the space of operators
$\partial_x - V_2$ with periodic potential which satisfy the constrains
$H_k=const,\; k=1,\ldots,n$; see for details \cite{KP}.

Before proceeding to the proof of the  Lemma, we compute the first two symplectic structures using the formula
$$
\omega_n=\sum\limits_{P_{\pm}} \R \frac{\l^n}{ \F(Q)}
<\e^+ J\delta V \wedge \delta \e> d \lambda,\qquad\qquad\qquad n=1,2.
$$
The result is
\begin{eqnarray}
\omega_0&=& 2i <\d \psib \wedge \d \psi>, \label{sy0}\\
\omega_1& =&  \omega_1= <\d \psi \wedge \d \psib' + \d \psib \wedge \d \psi' +
2\d \partial^{-1} |\psi|^2\wedge \d |\psi|^2>.\label{sy1}
\end{eqnarray}
subject to the constraint $H_1=const$.
We present the computaion divided in small steps.

\noi
{\it Step 1.} Identity $\F(\tau_{\pm}Q)=-\F(Q)$ implies
\begin{eqnarray*}
\frac{1}{ \F(Q)} & =\;\; \phi_0 + \frac{\phi_1}{ \l} + \frac{\phi_2}{ \l^2} +\cdots,
\quad \quad \quad \quad
Q\in (P_+), \\
\frac{1}{ \F(Q)} & =-\phi_0 - \frac{\phi_1}{ \l} - \frac{\phi_2}{ \l^2} - \cdots,
\quad \quad \quad \quad
Q\in (P_-).
\end{eqnarray*}
Using definition of $\e(x,Q)$ and $\e^+(x,Q)$ from previous section,  we have
\begin{eqnarray*}
\phi_0& = &1,\\
\phi_1& = & -<d_1+ \overline{d}_1>.\\
\end{eqnarray*}

\noi
{\it Step 2.} Near $P_+$ we obtain
$$
<\e^+ J\d V \wedge \d \e> = \frac{s_1}{ \l} + \frac{s_2}{ \l^2} +
 \frac{s_3}{ \l^3}  + \cdots,
$$
where
\begin{eqnarray*}
s_1& = &<\hat \e_0^T J \d V \wedge \d \e_1>,\\
s_2& = & <\hat \e_0^T J \d V \wedge \d \e_2> +
<\hat \e_1^T J \d V\wedge \d \e_1>.
\end{eqnarray*}
Similar, at $P_-$, we obtain
$$
<\e^+ J\d V \wedge \d \e> = - \frac{\bar{s}_1}{ \l} - \frac{\bar{s}_2}{ \l^2} -
 \frac{\bar{s}_3}{ \l^3}  - \cdots.
$$
Using the expansion for $\F(Q)$ from Step 1, we derive
\begin{eqnarray*}
\omega_0& = & \phi_0 (s_1 + \bar{s}_1),\\
\omega_1& = & \phi_0 (s_2 + \bar{s}_2) + \phi_1 (s_1 + \bar{s}_1).
\end{eqnarray*}
\noi
{\it Step 3.} Computing $s_1$, we have
$$
s_1=<\hat \e_0^T J \d V\wedge \d \e_1> =<\d \psi \wedge \d b_1>.
$$
Using the formula $b_1=-i \psib$, we obtain $s_1= -i <\d \psi \wedge \d \psib> $
and \ref{sy0}.

\noi
{\it Step 4.} The first term in the formula for $s_2$ produces
$$
<\hat \e_0^T J\d V \wedge \d \e_2>=< \d \psi \wedge \d b_2> .
$$
Using recurrence relation \ref{fir}: $ b_2=ib_1' -i \psib d_1 $, we obtain
$$
< \d \psi \wedge i \d b_1' - i \d \psib d_1 -  i \psib \d d_1>.
$$
The second term in the formula for $s_2$ produces
$$
<\hat \e_1^T J \d V \wedge \d \e_1> = <- \overline{b}_1 \d \psib \wedge \d d_1 +
\overline{d}_1 \d \psi \wedge \d b_1>.
$$
Finally,
$$
s_2= <\d \psi \wedge \d \psib' -i(d_1+\overline{d}_1) \d \psi \wedge \d \psib - i
\d |\psi|^2 \wedge \d d_1>.
$$
Using the formula for $s_1$, 
$$
\omega_1= <\d \psi \wedge \d \psib' + \d \psib\wedge \d \psi' - i \delta |\psi |^2\wedge
\d(d_1-\overline{d}_1)>.
$$
The constraint $H_1=const$ implies $<\d |\psi|^2>=0$ and using the explicit formula for
$d_1$, we obtain \ref{sy1}.

\noi
{\it Proof.}  Closeness of the form $\omega_0$ follows from the result of  next Lemma or from explicit formula \ref{sy0}. For the second statement 
we present  a complete proof only for $m=0$. Higher flows can be treated similarly. 
In order to prove that the first flow is Hamiltonian one has to establish, \cite{A}:
$$
i_{\partial_t} \omega_0= - \d\, 2H_1,
$$
where $i_{\partial_t}$ is the contraction operator produced by the vector field $X_1$.
Using time--dependent BA functions $i_{\partial_t} \delta \e = \e^{\bullet},\;\; i_{\partial_t} \delta V= V^{\bullet}$
we have   
$$
i_{\partial_t} \omega_0=  \sum\limits_{P_{\pm}} \R < \e^* J V^{\bullet} \delta \e> d\l -
\R <\e^* J \delta V \e^{\bullet}>d\l.
$$

Let us compute the residue at $P_+$.
From the computation preceding the proof 
\begin{eqnarray*}
\frac{1}{ \F(Q)} & = & \;\; \phi_0 + \frac{\phi_1}{ \l} + \ldots,
\quad \quad \quad \quad
Q\in (P_+), \\
\frac{1}{ \F(Q)} & = & -\phi_0 - \frac{\phi_1}{ \l} -  \ldots,
\quad \quad \quad \quad
Q\in (P_-).
\end{eqnarray*}
Then, using $V^{\bullet}=\[ \frac{i }{ 2} \sigma_3, V\]=i \sigma_3 V$, we obtain
$$
\R < \e^* J V^{\bullet} \delta \e> d\lambda  =
i \phi_0  <\hat \e_0^T J \sigma_3 Y_0 \delta \e_1>.
$$
Similarly, using $\e^{\bullet} = \frac{i}{ 2} \sigma_3  \e$  we have:
\begin{eqnarray*}
  \R   < \e^* J \delta V \e^{\bullet}> d\l
          & = &{i}{ 2 } \phi_1    <\hat{\e}_0^T J \d Y_0 \sigma_3 \e_0> \\
          & + & {i}{ 2}\phi_0 \[ < \hat {\e}_0^T J \delta Y_0 \sigma_3  \e_1 > +
                              < \hat {\e}_1^T J \delta Y_0 \sigma_3 \e_0 >\].
\end{eqnarray*}
The first term vanishes and
\begin{eqnarray*}
i_{\partial_t}   \R  <\e^* J\delta V \wedge \delta \e>d \lambda  &=& 
 i  \phi_0  <\hat{\e}_0^T J \sigma_3 Y_0  \delta \e_1> \\
& -  &\frac{i }{ 2 } \phi_0  \[  <\hat{\e}_0^T J \d Y_0 \sigma_3  \e_1> +   
< \hat{\e}_1^T J \d Y_0 \sigma_3  \e_0 >\].
\end{eqnarray*}

Similarly at $P_-$
\begin{eqnarray*}
i_{\partial_t}  \R  <\e^* J \delta V \wedge \delta \e> d \lambda  &=&
-i  \phi_0  <\e_0^T J \sigma_3 Y_0 \d  \hat{\e}_1> \\
& +&   \frac{i}{ 2 } \phi_0 \[ <\e_0^T J \d Y_0 \sigma_3 \hat{\e}_1> +
< \e_1^T J \d Y_0 \sigma_3 \hat{\e}_0 >\].
\end{eqnarray*}
Finally, we obtain
$$
i_{\partial_t} \omega_0 = i \phi_0 \[ <\hat{\e}_0^T J \sigma_3 Y_0 \d  {\e}_1> -
 <\e_0^T J \sigma_3 Y_0  \d \hat{\e}_1>\] =- \d \, 2H_1.
$$
\qed

\subsection{Darboux coordinates}

The formulas \ref{sy0}--\ref{sy1} give examples of symplectic forms. All these forms can be put in the 
Darboux form in the coordinates associated with  poles  of the Baker--Akhiezer function. 
\begin{lem} The formula
$$
\xi_0(Q)=<\e^* J\d V \wedge \d \e> d \l
$$
defines meromorphic in $Q$ differential 2-form on $\Gamma$ with poles
at $\gamma_1,\ldots, \gamma_{g+1}$ and $P_{+}, P_{-}$.
The symplectic 2-form defined by the formula
$$
\omega_0= \sum\limits_{P_{\pm}} \R\;  \xi_0(Q)
$$
can be written as
$$
\omega_0= \frac{2}{i} \sum\limits_{k=1}^{g+1} \d p(\gamma_k) \wedge \d \l (\gamma_k).
$$
\end{lem}
\noi
{\it Remark 1.} The meaning of the right-hand side of this formula is the following. The curve $\Gamma$ (or its cover $\hat{\Gamma}$) is equipped with two meromorphic functions $\l(Q)$ and $p(Q)$.  Their variation $\delta p(Q)$ and $\delta \l(Q)$ at the points of the divisor is computed for variation of the potential  $\psi(x),\; \psib(x);\, -l\leq x\leq l$. The RHS of the formula is the sum of an exterior products of these variations.

\noi
{\it Remark 2.} In fact for a general smooth potential the divisor $\l(\g_k)$ and  values of the quasimomentum $\cosh^{-1} \Delta(\l(\g_k))$ with 
suitably chosen sign (=sheet) 
determine the potential. In other words they are global coordinates on the phase space. First note that the discriminant $\Delta$  can be reconstructed from this data using Shannon interpolation, see \cite{MV}. Thus the curve $\G$ is known. The  potential can be effectively recovered from the divisor  via trace formulas, see \cite{MC2}.

\noi
{\it Proof.}  Note $\d V= \d Y_0$ does not
depend on $\l$. Essential singularity of the Floquet solutions at $P_{\pm}$ cancels out, and infinities are   simple poles for the form 
$\xi_0(Q)$. In the finite  part of the curve $\xi_0(Q)$ has two sets of poles. One is the poles  $\gamma_1,\ldots, \gamma_{g+1}$ of the Baker-Akhiezer function. Another is the branch points   of the curve $\G$. 
By the Cauchy theorem
$$
\sum\limits_{P_{\pm}} \R \; \xi_0(Q) =- \sum\limits_{\gamma_k}  \R\;  \xi_0(Q)- \sum\limits_{s_k}  \R\;  \xi_0(Q).
$$
 
Let us compute contribution of the first set of poles. 
Near $\gamma_k$ we have
$$
\e=\frac{\R \; \e}{\l-\l(\gamma_k)} +O(1).
$$
Therefore, 
$$
\d \e=  \frac{\R \,\e}{(\l - \l(\gamma_k))^2} \d \l(\gamma_k) +O\(\frac{1}{
\l-\l(\gamma_k)}\),
$$
and 
\beq\label{var}
 \d \e=  \frac{\e }{ \l - \l(\gamma_k)} \delta\l(\gamma_k) + O\(\frac{1}{ \l- \l(\gamma_k)}\).
\eeq
Note that $\e^*(x,\gamma_k)\equiv 0$ and from Lemma 2.3 we obtain
$$
 \R_{\gamma_k}\; \xi_0(Q)=  <\e^* J\d V \e> (\gamma_k)\wedge \, \d \l(\gamma_k) \;  \R_{\gamma_k}\;\[ \frac{d \l }{ \l - \l(\gamma_k)}\]= 
i \d p(\gamma_k)\wedge \d\l (\gamma_k).
$$

Now consider branch points  of the curve. They produce a nontrivial contribution, though the pole of $\e^*$ at the
branch point  is  annihilated by the zero of the differential $d\l$. Nevertheless, the variation $\delta \e(x,Q)$ has a simple pole at $s_k$. 
First, let us make a  general remark. 

Consider, the variation   of a function $f(Q,\psi,\psib)$ under variation of the potential  $\psi(x),\; \psib(x),\, -l\leq x\leq l;$ taken for $Q$  in the vicinity of the branch point $s_k$  and  a fixed value of $\l$. Such variation  will have a pole at the  branch point  itself.  At the branch point $\l$ fails to be a local parameter, but $w$ is fine due to the fact $\partial_{\l} R(\l,w)|_{s_k}\neq 0$.  Now, consider a function $f(Q,\psi,\psib)=f(w,\psi,\psib)$, and define its variation $\delta_0$ for a fixed value of $w$. Then, 
$$
\delta f=\delta_0 f + \frac{d f}{d w}\; \delta w.
$$   
Take, for example $f(Q)=\l(Q)$, then
\beq\label{so}
0=\delta_0 \l +\frac{d\l}{d w} \;\delta w.
\eeq
Therefore, for a general $f$ we have 
\beq\label{sov}
\delta f=\delta_0 f -\frac{d f}{d \l}\; \delta_0 \l=-\frac{d f}{d \l}\; \delta_0 \l+O(1).
\eeq
The zero of the differential $d\l$ at the branch point produces the pole of $\delta f$.

We can proceed to the computation of the residues of $\xi_0(Q)$ at $s_k$. 
In the local parameter
$(\l-\l(s_k))^{1/2}\sim w-1,\;\l=\l(Q)$:
$$
\delta\e(x,Q)=-\frac{\e_1(x)}{2}\frac{\delta\l(s_k)}{(\l-\l(s_k))^{1/2}}+\hdots,
$$
where
$$
\e(x,Q)=\e_0(x)+\e_1(x)(\l-\l(s_k))^{1/2}+\hdots,\qquad\qquad Q\in(s_k).
$$
Similarly, 
$$
d\e(x,Q)=\frac{\e_1(x)}{2}\frac{d\l}{(\l-\l(s_k))^{1/2}}+\hdots,
$$
and we have 
$$
\delta\e(x,Q)=-\frac{d\e(x,Q)}{d\l} \d\l(s_k)+O(1).
$$
The leading term is the same as in general formula \ref{sov}. Therefore, \ref{so} implies 
$$
\R_{s_k}\; \xi_0(Q)=-\R_{s_k}\[<\e^* J\d V d\e>\]\wedge \, \d\l(s_k)=\R_{s_k}\[<\e^* J\d V d\e>\wedge \, \frac{d\l\;\d w}{d w}\] .
$$

Now using  
\bey
\e^*(x,Q)&=&\e^*(l,Q)M^T(l,x,\l)^{-1},\\
d\e(x,Q)&=&dM(x,-l,\l)\e(-l,Q)+M(x,-l,\l)d\e(-l,Q)
\eey
with the help $dM(x,y,\l)|_{s_k}=0$     we obtain,
$$
\R_{s_k}\; \xi_0(Q)=\R_{s_k}\[\e^*(l,Q)<M^T(l,x,\l)^{-1} J\d V M(x,-l,\l)>d\e(-l,Q)\wedge \, \frac{d\l\;\d w}{d w}\] .
$$
Symmetry \ref{nsy} of the monodromy matrix implies
$$
<M^T(l,x,\l)^{-1} J\d V M(x,-l,\l)>=\frac{1}{2l}J\d M(l,-l,\l),
$$
and using skew-symmetry of the wedge product 
\bey
\R_{s_k}\; \xi_0(Q)&=&\frac{1}{2l}\R_{s_k}\[\e^*(l,Q)J\d M(l,-l,\l)d\e(-l,Q)\wedge \, \frac{d\l\;\d w}{d w}\]\\
&=&\frac{1}{2l}\R_{s_k}\[\e^*(l,Q)J(\d M(l,-l,\l)-\d w)d\e(-l,Q)\wedge \, \frac{d\l\;\d w}{d w}\].
\eey
Identities \ref{mp} and \ref{mps} imply 
\bey
\e^*J(\d M-\d w)&=&\d\e^*J(w- M),\\
J( w-M)d\e&=&J( dM- dw)\e.
\eey
Therefore,
$$
\R_{s_k}\; \xi_0(Q)=\frac{1}{2l}\R_{s_k}\[\d\e^*(l)J(dM(l,-l,\l)-dw)\e(-l)\wedge\; \frac{d\l\;\d w}{d w}\] .
$$
Since $\e^*(l)J\e(-l)=w^{-1}$, we have 
$$
\R_{s_k}\; \xi_0(Q)=\frac{1}{2l}\R_{s_k}\[\e^*(l)J\d \e(-l)\wedge\,\d w\, d\l\].
$$
The one form  
$$
\e^*(l,Q)J\d \e(-l,Q)\wedge\,\d w(Q)\, d\l(Q)
$$
is holomorphic (in the parameter $\l$) outside of the poles $\g_k$ and the branch points. At infinity the essential singularity cancels out and 
due to \ref{asf}-\ref{assf}, Lemmas \ref{lfloc} and  \ref{lsfloq}
$$
e^*(l,Q)J\d \e(-l,Q)\wedge\,\d w(Q)=o\(\frac{1}{\l}\).
$$
This implies
$$
\R_{P_{\pm}}\[\e^*(l,Q)J\d \e(-l,Q)\wedge\,\d w(Q)\, d\l(Q)\] =0.
$$ 
By the Cauchy theorem, 
$$
\sum_{s_k}\R_{s_k}\[\e^*(l)J\d \e(-l)\wedge\; d\l\;\d w\]=-\sum_{\gamma_k}\R_{\gamma_k}\[\e^*(l)J\d \e(-l)\wedge\; d\l\;\d w\]
$$
Therefore, using $\e^*(l)=w^{-1}\e^*(-l)$, we have 
\bey
\sum_{s_k}\R_{s_k}\; \xi_0(Q)&=&-\frac{1}{2l}\sum_{\gamma_k}\R_{\gamma_k}\[\e^*(l)J\d \e(-l)\wedge\; d\l\;\d w\]\\
&=&-\frac{1}{2l}\sum_{\gamma_k}\R_{\gamma_k}\[\e^*(-l)J\d \e(-l)\wedge\; d\l\;\frac{\d w}{w}\]
\eey
Using the formula \ref{var} we finally obtain 
\bey
\sum_{s_k}\R_{s_k}\; \xi_0(Q)&=&\sum_{\gamma_k}i\d p(\gamma_k)\wedge \R_{\gamma_k}\[\e^*(-l)J\d \e(-l) d\l\]\\
&=&\sum_{\gamma_k}i\d p(\gamma_k)\wedge \d \l(\gamma_k)\R_{\gamma_k}\[ \frac{d\l}{\l-\l(\gamma_k)}\]\\
&=&\sum_{\gamma_k}i\d p(\gamma_k)\wedge \d \l(\gamma_k).
\eey
\qed

\noi
{\it Remark.}  For the higher symplectic structures  an analogous result holds
$$
\omega_n=\frac{2}{i} \sum\limits_{k=1}^{g+1} \l^n \d p(\gamma_k) \wedge \d \l (\gamma_k),
\quad \quad \quad \quad n=0, 1 \ldots;
$$
subject to the constrains
$H_k=const,\; k=1,\ldots,n$.
\subsection{Action--angle variables}
Here we describe briefly another system of Darboux coordinates. We refer to the paper \cite{MV} for details. 

The actions $I_k,\;k=1,\hdots, g+1;$ are defined by the formula 
$$
I_k=\frac{1}{4\pi} \int_{a_k} p(\l) d\l.
$$
In the formula above the multivalued function $p(\l)$ is normalized such that $p(s_k)=0,\;k=1,\hdots, g+1$. 
The angles  $\t_k,\;n=1,\hdots, g+1;$ are
$$
\t_k=\sum_{n=1}^{g+1} \int_{s_n}^{\gamma_n} \alpha_k.
$$
The differentials $\alpha_k,\;n=1,\hdots, g+1;$ are of the third kind with poles at the infinities $P_{\pm}$  
normalized in such a way that\footnote{$\d_n^k$ is  Kronecker delta.} 
$$
\int_{a_k}\alpha_n=2 \pi \d_n^k.
$$
As it is proved in \cite{MV} by a direct computation
$$
\{\t_n,I_k\}=\d_n^k.
$$ 
All other brackets vanish 
$$
\{I_n,I_k\}=0,\qquad\qquad \{\t_n,\t_k\}=0.
$$
These formulas imply the identity for symplectic forms.
$$
\omega_0=2 \sum_{n=1}^{g+1} \d I_n\wedge \d \t_n.
$$

We conclude this section with a few  remarks.

\noi 
{\it  Remark 1.}
Another way to prove the identity for symplectic forms without employing the Poisson bracket is found by  Krichever, \cite{K2}. 

\noi 
{\it Remark 2.} McKean, \cite{MC1}, proved various identities   for  1--forms.

\noi
{\it  Remark 3.} In the finite--gap case the curve $\G$ is specified  by $2(g+1)$ branch points. 
The actions  $I_k,\;k=1,\hdots,g+1;$ together with the $g$ other periods (see \ref{pc}) of the differential $dp$ and the 
constant $p_0^{+}$ determine  the curve. This fact  is due to Krichever, \cite{K3} (see  also \cite{BK}).
In the infinite--gap case  it is shown in \cite{MV} that $I-\t$'s are global coordinates on the phase space for a general square integrable potential.  
This property holds for a finite gap potential as well.

\section{The Scattering Case}
\subsection{Jost solutions}
In the next  sections we consider the scattering  problem for the Dirac operator  on the entire line with 
rapidly decaying potential.   
The Riemann-Hilbert approach to the  scattering theory for  canonical systems 
with summable potential was constructed by M.G. Krein and P.E. Melik--Adamian,
\cite{K, KMA, MA}. This approach was used many times  in soliton theory, \cite{FT}.

To simplify the estimates we assume that the potential $\psi$ is from the Schwartz'
space $S(\RB)$ of complex rapidly decreasing infinitely differentiable functions on the  line such that
$$
\sup_{x} |(1+x^2)^n \, \psi^{(m)}(x)|\, < \infty\qquad\qquad\qquad\qquad m,\,n=0,1,\ldots.
$$
Let us introduce the reduced transition  matrix $T(x,y,\l),\, x\geq y;$ by the formula
\bay\label{red}
T(x,y,\l)=E^{-1}\(\frac{\l x}{ 2}\) M(x,y,\l) E^{-1}\(-\frac{\l y}{ 2}\),
\ey
where $E(\frac{\l x}{ 2})= \exp{(-\frac{i\lambda x}{ 2}  \sigma_3)}$ is a solution of
the free equation ($\psi\equiv 0$).  The matrix $T(x,y,\l)$ solves the equation
$$
T'(x,y,\l) = Y_0(x) E(\l x) T(x,y,\l), \qquad\qquad
\quad T(y,y,\lambda)= I.
$$
The spectral parameter enters multiplicatively into the RHS of the differential equation.
The solution is given by the formula
\bay\label{exp}
T(x,y,\lambda)= \exp \, \int_{y}^x Y_0(\xi) E(\l\xi) d\xi. 
\ey
 The symmetry of the matrix
$Y_0:\, \sigma_1 Y_0(x)\sigma_1= \overline{Y_0(x)}$
is inherited by unimodular matrix 
T:$$\; \sigma_1T(x,y,\overline{\lambda})\sigma_1=
\overline{T(x,y,\lambda)}.$$
For  real  $\l$ the formula \ref{exp} and the rapid decay of the potential imply an existence of the limit
$$
T(\lambda)= \lim T(x,y,\lambda)= \(\begin{array}{ccccc}
a(\lambda) & \overline{b}(\lambda)\\
b(\lambda) & \overline{a}(\lambda) \end{array} \), \quad
{\rm when} \quad y \rightarrow -\infty \quad\text{and}\quad x
\rightarrow +\infty;
$$
and  $|a(\lambda)|^2- |b(\lambda)|^2=1.$ 
When  the potential $\psi\in  S(\RB)$, we have   $b(\l) \in S(\RB)$.

We introduce {\it Jost solutions}  $J_{\pm}(x,\lambda)$ as a matrix solutions of the differential equation
$$
J_{\pm}'(x,\lambda)=V(x,\lambda)J_{\pm}(x,\lambda),\quad J_{\pm}(x,\l)=
E\(\frac{\lambda x}{ 2}\) + o(1),\quad {\rm when}\quad x\rightarrow \pm \infty.
$$
An existence and analytic properties of the Jost solutions follow from the integral
representations
$$
J_+(x,\lambda)=E\(\frac{x\lambda}{ 2}\)+ \int_x^{+\infty} \Gamma_+(x,\xi)
E\(\frac{\lambda \xi}{ 2}\) d\xi,
$$
$$
J_-(x,\lambda)=E\(\frac{x\lambda}{ 2}\)+ \int_{-\infty}^x \Gamma_-(x,\xi)
E\(\frac{\lambda \xi}{ 2}\) d\xi.
$$
The kernels $\Gamma_{\pm}$ are unique and infinitely smooth in both variables.
Introducing the   notation $J_\pm=\[\jo^{(1)}_\pm,\, \jo^{(2)}_\pm\]$ we see from the
integral representations that
$\jo^{(1)}_-(x,\l),$ $  \jo^{(2)}_+(x,\l)$ are analytic in $\l$ in the upper half-plane
and continuous up to the boundary. Also, the columns
$\jo^{(2)}_-(x,\l),\;\; \jo^{(1)}_+(x,\l)$
are analytic  in the lower half--plane and continuous up to the boundary.

Now we describe analytic properties of the coefficient $a(\l)$  of the matrix $T(\l)$. 
The monodromy matrix $M(x,y,\l)$ can be written in the form
$$
M(x,y, \l)=J_+(x)J^{-1}_{+}(y)= J_{-}(x)J^{-1}_{-}(y).
$$
Therefore, 
$$
J_{+}^{-1}(x)\, M(x,y,\l)\, J_{-}(y)= J^{-1}_{+}(y)J_{-}(y)=J_{+}^{-1}(x)J_{-}(x).
$$
The variables $x$ and $y$ separate and  the above expression does not depend on $x$ or $y$ at all. 
By passing to the limit  with
$ x\rightarrow +\infty,\, y \rightarrow -\infty$ we have 
$$
T(\l)=J^{-1}_{+}(y)J_{-}(y)=J_{+}^{-1}(x)J_{-}(x).
$$
Therefore,  
$$a(\l)= {\jo_-^{(1)}}^T(\l)J \jo_+^{(2)}(\l).
$$
The properties of Jost solutions imply that 
\begin{itemize}
\item  $a(\lambda)$ is analytic in the upper half-plane and continuous  up to the boundary;

\item   $a(\lambda)$ is  root-free;
\item $|a(\lambda)| \geq 1$ 
and $|a(\lambda)|^2-1 \in S(\RB)$ for $\l$ real, $a(\lambda) = 1 +o(1)$ as $ |\lambda| \longrightarrow \infty$. 
\end{itemize}
This coefficient will be used to construct the scattering curve $\Gin$. 

Let $p_\infty(\lambda)$ be  such that $a(\lambda)= \exp(-i2 p_{\infty}(\lambda))$
for $\lambda$ in the upper half-plane. The quantity $\pin$ in analogous to the quasimomentum studied in  the periodic case, see \cite{V1}\footnote{The
spectral parameter $\l$ in the paper \cite{V1} has different scaling and should
be replaced by $\lt$.}. From the properties of $a(\lambda)$
it follow that $ \pin $  is analytic in the upper half-plane and
continuous  up to the boundary;  $\Im \pin \geq 0$ for $ \Im \lambda \geq  0$; 
$\pin=o(1)$ for $|\lambda| \rightarrow \infty$;
for real $\lambda$,  the density of the measure $d \mi= \Im \pin \, d \lambda$ belongs to  
$S(\RB)$.  The function $\pin$ can be written in the form
$$
\pin=\frac{1}{ \pi}\int \frac{d \mu_{\infty}(t)}{ t -\lambda}.
$$
Expanding the denominator in inverse powers of $\lambda$, we obtain:
\bay\label{V}
\pin= -\sum^{\infty}_{k=0}\frac{1}{ \lambda^{k+1}} \frac{1}{ \pi}\int_{-\infty}^{+
\infty} t^k d \mu_{\infty}(t)= -\frac{ H_1}{ \lambda}-  \frac{ H_2}{ \lambda^{2}}-
\frac{ H_3}{  \lambda^3}+ \dots .
\ey
where $H_1,H_2$ and $H_3$ are the integrals introduced above with $l=+\infty$.
The expansion has an asymptotic character for $\lambda: \; \delta \leq
\arg \lambda \leq \pi -  \delta,\;\delta > 0$.

To  describe the asymptotic behavior in $x$  of the Jost solutions  $\jo_+^{(2)}(x,\l)$ and $\jo_-^{(1)}(x,\l)$ we assume that $\l$ is real and fixed. Then, 
$$
\begin{array}{ccccccccccccccccc}
&\phantom{kkkkkkkkk} &x \rightarrow - \infty  &  &x\rightarrow  +  \infty \\
&\jo_+^{(2)}   &a(\l) \f_{\rightarrow}(x,\l)-\bar{b}(\l)\f_{\leftarrow}(x,\l)&   &
\quad \f_{\rightarrow}(x,\l)\\
&\jo_-^{(1)}   &\f_{\leftarrow}(x,\l) &\phantom{h} &
a(\l)\f_{\leftarrow}(x,\l)+
b(\l) \f_{\rightarrow}(x,\l),&
\end{array}
$$
where
$$
\f_{\leftarrow}(x,\l)=\left[ \begin{array}{ccccccccccccccccc}
                                           e^{- i\lt x}\\
                              0\end{array} \right], \quad
\f_{\rightarrow}(x,\l)=\left[ \begin{array}{ccccccccccccccccc}  0\\
                    e^{i\lt x}\end{array} \right]
$$ are  solutions of the free equation.

We sketch the derivation of the asymptotics for  $\jo^{(2)}_+(x,\l)$, when  $x\rightarrow -\infty$.
Let $\psi$ be a potential that vanishes outside the segment $[-L, +L]$.
In this case  formula \ref{exp} becomes 
$$
T(\lambda)= \exp
\, \int_{-L}^L Y_0(\xi) E(\l\xi) d\xi,
$$
and $T(\l)$ is an entire unimodular function of $\l$ of the form
$$
T(\l)= \[\begin{array}{cccccccc}  a(\l) &   \overline{b}(\overline{\l})\\
                 b(\l) &  \overline{a}(\overline{\l}) \end{array} \].
$$
From the definition of the  matrix $M$:
\bay\label{for}
\jo^{(2)}_+(L,\l)=M(L,-L,\l) \jo^{(2)}_+(-L,\l),
\ey
and from \ref{red} 
$$
M(L,-L,\l)=\[\begin{array}{cccccccc} a(\l) e^{-i\l L} & \overline{b}(\overline{\l}) \\
                     b(\l)    &       \overline{a}(\overline{\l}) e^{i\l L} \end{array} \].
$$
Obviously, $\jo^{(2)}_+(L,\l)= \fr(L,\l)$ and 
$\jo^{(2)}_+ (-L,\l)=c_1 \fr(-L,\l) + c_2 \fl (-L,\l)$ with unknown coefficients $c_1$ and $c_2$. Formula \ref{for} for real  $\l$ 
leads to the linear system 
$$
\begin{array}{cccccccc}  
c_1 \overline{b}({\l}) + c_2 a(\l) & =0\\
c_1 \overline{a}({\l}) + c_2 b(\l)&  =1.
\end{array}
$$
Solving for $c$'s we obtain the stated formula. For a potential with noncompact
support one has to take $L$ sufficiently large to make the error negligible.

The Riemann surface $\Gin$ is obtained by gluing together  along the real line  
two copies of the complex plane (see Figure 2). One copy we call "+" and another "-". 
Each copy has an infinity $P_{+}$ or $P_{-}$. The point $Q\in \Gin$ is determined by $\l=\l(Q)$ 
and specification of the sheet $Q=(\l,\pm)$.  Let us define  for the "+"  copy 
$\jo(x,Q)$ to be   $\jo^{(2)}_+(x,\l)$ if $ \Im \l > 0;$  and  
$\jo^{(2)}_-(x,\l)$ if $\Im \l <  0$. For the "-" copy we define $\jo(x,Q)$  to be 
$\jo^{(1)}_-(x,\l)$ if $\Im \l > 0;$ and $\jo^{(1)}_+(x,\l)$ if $\Im \l < 0$.
In the vicinity of  $P_{\pm}$ the  function $\jo(x,Q)$ has  asymptotics, 
\bay\label{J}
\jo(x,Q) =  e^{\pm i\frac{\l}{ 2}x} \[\jo_0/\hat \jo_0  + o\(1\)\],  
\ey
where $ \l=\l(Q)$ and 
$$
\jo_0= \[\begin{array}{ccc} 0\\1\end{array}\]\qquad\qquad\qquad 
\hat \jo_0= \[\begin{array}{ccc} 1\\0\end{array}\].
$$
Therefore, $\jo(x,Q)$ can be viewed as a BA function for the singular curve $\Gin$.

We also introduce   the matrix BA function  
$$
H_+(\l)=\[ \jo_-^{(1)}(\l),\jo_+^{(2)}(\l)\] \quad {\rm and}\quad\quad H_-(\l)=\[ \jo_+^{(1)}(\l),\jo_-^{(2)} (\l) \]
$$ analytic in the upper/lower half-plane respectively. They are connected by the gluing condition
\bay\label{RH}
H_-(x, \l)= H_+(x,\l)  S(\l),\qquad\qquad\qquad\text{where}\qquad\l \in \RB
\ey
and the scattering matrix $S(\l)$
$$
S(\l) = \frac{1}{ a} \[\begin{array}{ccccc} \;\; {1} & {\overline{b}}\\
                       -{b}& {1}
                          \end{array}\].
$$

The  adjoint (dual) Jost solution $\jo^+$ at the point $Q$ is defined by the formula
$$
\jo^+(x,Q)\equiv \jo(x,Q)^T.
$$
Any Jost solution satisfies $\[J\partial_x- J V\]\jo=0$.
By analogy with  the periodic case  one can prove that $\jo^+$ satisfies
$\jo^+\[J\partial_x- J V\]=0.$

The matrices  $H_+^+$ and $H_-^+$ are  defined as
$$
H_+^+(\l)=\sigma_1 H_+^T(\l)= \[\begin{array}{ccccccc}   \jo_+^{(2)\,T}\\
                                     \jo_-^{(1)\,T}
              \end{array}\],  \qquad\qquad 
H_-^+(\l)=\sigma_1 H_-^T(\l)= \[\begin{array}{ccccccc}   \jo_-^{(2)\,T}\\
                                     \jo_+^{(1)\,T}
\end{array}\]. 
$$
Extending $a$ into the lower half-plane by the formula  $a^*(\l)= \overline{a(\bar \l)}$, we define
$$
H_+^*(\l)= -\frac{\sigma_3}{ a(\l)} H_+^+(\l)\qquad \qquad \qquad\qquad \text{for}\qquad \Im \l >0;
$$
and
$$
\qquad  H_-^*(\l)=  -\frac{\sigma_3}{ a^*(\l)} H_-^+(\l)\qquad \qquad \qquad \text{for}\qquad \Im \l < 0.
$$
It is easy to check that the dual  gluing condition holds
\bay\label{ARH}
H_-^*(x,\l) =S^{-1}(\l) H_+^*(x,\l),\qquad\qquad\qquad{\rm where} \qquad \l\in \RB
\ey
and
$$ S^{-1}(\l)= \frac{1}{ a^*} 
\[\begin{array}{ccccc} \;\; {1} & -{\overline{b}}\\
                        \;\;{b}& {1}
    \end{array}\].
$$

Next two lemmas state asymptotic properties of  Jost solutions which will be used in computations  with symplectic forms.
\begin{lem} (i) For  fixed $x$ the following formulas hold
$$
\jo_+^{(2)}(x,\l)=e^{+i \lt x}  \sum\limits_{s=0}^{\infty} \jo_s(x) \l^{-s} =
e^{+i \lt x} \sum\limits_{s=0}^{\infty} \[\begin{array}{cccc}
 g_s \\
 k_s
\end{array} \] \l^{-s}, 
$$
where $g_0=0, \quad k_0=1$, and\footnote{Operation $\hat{\phantom
0}$ applied to a scalar signifies complex
conjugation and reversal of infinities, see formulas in the  part (ii) below.}
$$
\jo_-^{(1)}(x,\l)=e^{-i \lt x} \sum\limits_{s=0}^{\infty} \hat \jo_s(x) \l^{-s}
= e^{-i \lt x} \sum\limits_{s=0}^{\infty} \[\begin{array}{cccc}
 \hat k_s \\
 \hat g_s
\end{array} \] \l^{-s} , 
$$
where $\hat g_0=0,\quad \hat k_0=1$. The expansion has an asymptotic character for
 $\l\;:\quad \d \leq \arg \l \leq \pi -\d,\;\d >0$.

(ii) The coefficients $g_1, \; k_1$ are given by the formulas
$$
g_1= -i \psib,\quad \quad \quad \quad
k_1= i \int_{x}^{+ \infty} |\psi(x')|^2 dx'
$$
and
$$
\hat g_1= i \psi, \quad \quad \quad \quad
\hat k_1 = i \int_{-\infty}^{x} |\psi(x')|^2 dx'.
$$
\end{lem}

{\it Proof} {\it (i).} Using\footnote{$\Gamma=\[\Gamma^{(1)},
\Gamma^{(2)}\]$.}    $\partial_\xi^n \Gamma_+^{(2)}(x,\xi)|_{\xi=\infty}= 0$, for
$n=0,1,...$ and integrating $n$ times by parts, 
\bay
\jo^{(2)}_+(x,\l)& = & e^{\frac{i\l x}{ 2}} \[\begin{array}{ccccc} 0\\ 1 \end{array} \]  +
\int_{x}^{\infty} \Gamma_+^{(2)} (x,\xi) e^{\frac{i\l \xi}{ 2}} d\xi  \nonumber  \\
 \dots      & =&  e^\frac{i\l x}{ 2} \[\begin{array}{ccccc} 0\\ 1 \end{array} \] -
\frac{  e^\frac{i\l x}{ 2} }{ \(\frac{i\l}{ 2}\)} \Gamma^{(2)}_+ (x,x)  +
\frac{  e^\frac{i\l x}{ 2} }{ \(\frac{i\l}{ 2}\)^2} \partial_\xi \Gamma^{(2)}_+ (x,x)
- \dots  \nonumber  \\
& & \quad +(-1)^n \frac{ e^\frac{i\l x}{ 2} }{\(\frac{i\l}{ 2}\)^n} \partial_\xi^{n-1} \Gamma^{(2)}_+ (x,x)
+ (-1)^n \frac{1}{ \(\frac{i\l}{ 2}\)^n} \int_{x}^{\infty} \partial_\xi^{n}
\Gamma_+^{(2)} (x,\xi) e^\frac{i\l \xi}{ 2} d\xi.  \nonumber 
\ey
This implies the  existence and   asymptotic character of the expansion in the parameter
$\l$. The other  infinity can be treated similarly.

{\it (ii).} Consider $\jo^{(2)}_+(x,\l)$ first.  The differential equation $ \jo'=V\jo$ implies
$$
-\[\begin{array}{ccccc}
g_s'\\
k_s'
\end{array}
 \] + Y_{0}
\[\begin{array}{ccccc}
g_s\\
k_s
\end{array}
 \]= \frac{i}{ 2} \(I +  \sigma_3 \)
\[\begin{array}{ccccc}
g_{s+1}\\
k_{s+1}
\end{array}
 \], \quad\quad s=0,1,\ldots;
$$
and $g_0=0,\quad k_0=1$.

This recurrent relation leads to the identities
\bay\nonumber
- g_s'+ \psib k_s &=& ig_{s+1}, \\
-  k_s' +\psi g_s &=& 0. \nonumber
\ey
For $s\geq 1$ we have the boundary condition
$$
g_s(x)|_{x=+ \infty}=k_s(x)|_{x=+ \infty}=0.
$$
These imply the stated  formulas for $g_1,\, k_1$.

For $\jo^{(1)}_-(x,\l)$  the differential equation  implies
$$
-\[\begin{array}{ccccc}
k_s'\\
g_s'
\end{array} \] + Y_{0}
\[\begin{array}{ccccc}
k_s\\
g_s
\end{array}
 \]= \frac{i}{ 2} \(- I +  \sigma_3 \)
\[\begin{array}{ccccc}
k_{s+1}\\
g_{s+1}
\end{array}
 \], \quad\quad s=0,1,\ldots;
$$
and $k_0=1,\quad g_0=0$.

The recurrent relation produces   the identities
\bay
- g_s'+ \psi k_s&= &- ig_{s+1}, \nonumber\\
- k_s' +\psib g_s&=&0. \nonumber
\ey
For $s\geq 1$ we have the boundary condition
$$
g_s(x)|_{x=- \infty}=k_s(x)|_{x=- \infty}=0.
$$
These imply the stated  formulas for $\hat g_1,\, \hat k_1$. We are done. \qed

\noi
{\it Remark.} It is interesting to compare asymptotic expansions for $\jo_-^{(1)}/ \jo_+^{(2)}$ and
$\e(x,Q)$. For the Jost solution $\jo_{-}^{(1)}(x,\l)$ normalized at the left 
$$
\jo_{-}^{(1)}(x,\l)= e^{ -\frac{i\l x}{ 2} } \( \[\begin{array}{ccccc} 1\\0 \end{array} \]  + \frac{1}{ \l}
\[ \begin{array}{ccccc} i\int_{-\infty}^x |\psi|^2\\ i\psi \end{array}
 \] + \dots \)
$$
and
$$
e^{i\lt l} \e(x,Q)= e^{- \frac{i\l x}{ 2}} \( \[\begin{array}{ccccc} 1\\0 \end{array}
 \]  + \frac{1}{ \l}
\[ \begin{array}{ccccc} -i \psi_0 + i\int_{-l}^x |\psi|^2\\ i\psi \end{array}
 \] + \dots \), \quad \quad Q\in(P_-).
$$
If $\psi$ is compactly supported and $l$ becomes sufficiently large, then
$e^{i\lt l} \e(x,Q)=\jo(x,\l)$.
For the Jost solution normalized at the right  the situation is slightly different. If one defines the {\it new}
$\e(x,Q)\equiv   \e(x,Q) w^{-1} (Q)$  which is the Floquet solution
normalized at the right end
$x=l$ of the interval, then for compactly supported $\psi$ and sufficiently large $l$
the {\it new} $e^{-i\lt l} \e(x,Q)= \jo_+^{(2)}(x,\l)$.

A result similar to Lemma 4.1 holds for Jost  solutions analytic in the lower half-plane. 
\begin{lem} (i) For  fixed $x$ the following formulas hold
$$
\jo_+^{(1)}(x,\l)=e^{-i \lt x}  \sum\limits_{s=0}^{\infty} \jo_s(x) \l^{-s} =
e^{-i \lt x} \sum\limits_{s=0}^{\infty} \[\begin{array}{cccc}
 h_s \\
 f_s
\end{array} \] \l^{-s},
$$
where $h_0=1, \quad f_0=0$, and
$$
\jo_-^{(2)}(x,\l)=e^{+i \lt x} \sum\limits_{s=0}^{\infty} \hat \jo_s(x) \l^{-s}
= e^{+i \lt x} \sum\limits_{s=0}^{\infty} \[\begin{array}{cccc}
 \hat f_s \\
 \hat h_s
\end{array} \] \l^{-s} ,
$$
where $\hat h_0=1,\quad \hat f_0=0$. The expansion has an asymptotic character for
 $ \l\;:\quad-  \d \geq \arg \l \geq -\pi + \d,\; \d >0$.

(ii) The coefficients $h_1, \; f_1$ are given by the formulas
$$
f_1= i \psi,\quad \quad \quad \quad
h_1= - i \int_{x}^{+ \infty} |\psi(x')|^2 dx'
$$
and
$$
\hat f_1= - i \psib, \quad \quad \quad \quad
\hat h_1 =-  i \int_{-\infty}^{x} |\psi(x')|^2 dx'.
$$
\end{lem}

Similar to the periodic case we will need time dependent BA functions. They are  obtained by an 
elementary construction.

\begin{lem} \cite{I}. 
There exists  the  Jost solution  $\jo(\tau,x,t, Q)$ on the curve $\Gin$ with three time parameters $\tau,\, x $ and $t$
which   satisfies the differential equations:
$$
[\partial_{\tau}-V_1(\tau,x,t)]  \jo(\tau,x,t, Q)  = 0,
$$
$$
[\partial_{x}-V_2(\tau,x,t)]   \jo(\tau,x,t, Q)  = 0,
$$
$$
[\partial_{t}-V_3(\tau,x,t)]  \jo(\tau,x,t, Q)  =0.
$$
\end{lem}

\noi 
{\it Proof.} 
We consider "+" sheet and the upper half-plane where $\jo(x,Q)=\jo_+^{(2)}(x,\l)$. 
First, we construct  $\jo(\tau,x,t, Q)$ such that
$
[\partial_{x}-V_2(\tau,x,t)] \;  \jo(\tau,x,t, Q)  =0
$
normalized for all $\tau$ and $t$ as
$$
\jo(\tau,x,t, Q) \sim e^{+i \lt x} \( \[\begin{array}{cccc}
 0  \\ 1  
\end{array} \] + o(1) \), \quad \quad \quad {\rm when} \quad x\rightarrow \infty.
$$

Then, we construct  $\jo(\tau,x,t, \epsilon_\pm Q)$ on the lower sheet  as a solution 
$$
[\partial_{x}-V_2(\tau,x,t)] \;  \jo(\tau,x,t, \epsilon_\pm Q)  =0
$$
for the same value of the spectral parameter $\l=\l(Q)$ normalized for all $\tau$ and $t$ as
$$
\jo(\tau,x,t, \epsilon_\pm Q) \sim e^{-i \lt x} \( \[\begin{array}{cccc}
 1  \\ 0 
\end{array} \] + o(1) \), \quad \quad \quad {\rm when} \quad x\rightarrow -\infty.
$$
The solutions $\jo(\tau,x,t,  Q)$ and $\jo(\tau,x,t, \epsilon_\pm Q)$ span the kernel of the operator $[\partial_{x}-V_2(\tau,x,t)]$. 
Now  we introduce
$$
\jo_{new}(\tau,x,t, Q)\equiv e^{-i\frac{1}{ 2} \tau - i \frac{\l^2}{ 2} t} \jo(\tau,x,t, Q),
$$
which is the desired solution.
Evidently,
$$
\[ \partial_{x} -V_2(\tau,x,t)\] \jo_{new}(\tau,x,t,Q)= 0.
$$

To prove the first identity of the statement we note, that 
commutativity of the operators $\partial_{\tau} - V_1$ and $\partial_{x} - V_2$ implies
$$ 
\[ \partial_{\tau} -V_1(\tau,x,t)\] \jo_{new}(\tau,x,t,Q)= c_1(\tau, t) \jo(\tau,x,t,Q) +
c_2(\tau, t) \jo(\tau,x,t,\epsilon_{\pm}Q).
$$
From another side as $x\rightarrow + \infty$,
$$
\[ \partial_{\tau} -V_1(\tau,x,t)\] \jo_{new}(\tau,x,t,Q)= \[\partial_{\tau}- \frac{i}{2}\sigma_3\]  e^{-i\frac{1}{ 2} \tau +
i \frac{\l}{ 2}x   - i \frac{\l^2}{ 2} t}  \( \[\begin{array}{cccc}
 0  \\   
 1
\end{array} \] + o(1) \) =o(1).
$$
Due to the linear independence of the solutions $ \jo(\tau,x,t, Q)$ and
$\jo(\tau,x,t,\epsilon_{\pm}Q)$ we have
$c_1(\tau, t) =c_2(\tau, t) =0$. 

Similarly  it can be proved that 
$$ 
\[ \partial_{t} -V_3(\tau,x,t)\] \jo_{new}(\tau,x,t,Q)=0.
$$
Another sheet of $\Gin$   can be treated  the same way. We are done. \qed

\noi
{\it Remark.}
It is easy to see that,
$$
\jo_{new}(\tau,x,t, Q) = e^{\pm i \( -\frac{1}{ 2} \tau + \frac{\l}{ 2} x - \frac{\l^2}{ 2}t\)}
\[ \jo_0/ \hat \jo_0 + o(1)\]\qquad\qquad\qquad Q\in (P_{\pm}).
$$
Thus the standard Jost solution with asymtotics \ref{J} can be obtained from the BA functions if 
one puts $\tau$ and $t$ equal to 0.

\subsection{ The symplectic structures } We are ready to introduce the scattering version of the  Krichever--Phong formula. 
The everaging is defined now as an integral over the entire line $<\bullet>=\int_{-\infty}^{+\infty}dx$. 

\begin{thm}  The formula
\bay\label{SY}
\omega_0=
\tr\,  \R\;   \frac{1}{ 2} \[<H_+^* J\delta V \wedge \delta H_+> + 
   <H_-^* J\delta V \wedge \delta H_->\]
d \lambda, 
\ey
defines a closed 2--form $\omega_0$ on the space of operators
$\partial_x - V_2$ with  potential from the Schwartz class $S(\RB)$.
The flows $e^{tX_m},\; m=1,2,\ldots$ on the space of operators defined by the
formula
$$
[\partial_{\tau_m} - V_m,  \partial_x -V_2]=0,
$$
are Hamiltonian  with respect to the  2--form  $\omega_0$ with  Hamiltonian
function $H_m$ (up to a non-essential constant factor). 
\end{thm}

\noi
{\it Remark 1.} The symbol 
$$
  \R\;    <H_+^* J\delta V \wedge \delta H_+>
$$
means the coefficient corresponding to the term $\frac{1}{\l}$ in the  power series expansion near infinity in the 
upper half--plane. The second term
$$
 \R\;    <H_-^* J\delta V \wedge \delta H_->
$$
is defined in the same way, only the upper half--plane plane is replaced with the lower half--plane.

\noi
{\it Remark 2.}  The formula
$$
\omega_n=\tr\;  \R\;   \frac{\l^n}{ 2}\[ <H_+^* J\delta V \wedge \delta H_+> + 
  <H_-^* J\delta V \wedge \delta H_->\]
  d \lambda,
$$
where $n=0, 1 \ldots,$ defines a closed 2--forms $\omega_n$ on the space of operators
$\partial_x - V_2$  with  potential from the Schwartz class $S(\RB)$ that satisfy the constraints
$H_k=const,\; k=1,\ldots,n$.
It is instructive to compute explicitly the symplectic forms for small $n$. 
The  first few    are given by the formulas
$$
\omega_0  =  2i< \d\overline{\psi}\wedge \d \psi>, 
$$
and 
$$
\omega_1  =  <\d \psi \wedge \d \psib' + \d \psib \wedge \d \psi' +
\delta \[\int_{-\infty}^x |\psi|^2 - \int^{\infty}_x |\psi|^2\]
\wedge \d |\psi|^2>. 
$$
subject to the constrain $H_1=const$. The derivation employs Lemmas 3.1-3.2 and similar to the 
periodic case.

\noi
{\it Proof. } Closeness of the form $\omega_0$ follows either from the explicit formula or from the result of the next theorem.  
The proof of the second statement we present  for the first $e^{tX_1}$ flow. The time dependent Jost solutions entering into \ref{TT} are constructed in Lemma 3.4. Let $i_{\partial_t}$ be the  construction  operator produced by the vector field $X_1$. We will prove
$i_{\partial_t}\omega_0=-\d 2H_1$.  For the first term in  \ref{SY}
\begin{eqnarray}
\tr & &\R\,    <  H_+^* J\delta V \wedge \delta H_+> \, d\l \nonumber \\
           & & = \R \frac{1}{ a} < \jo_-^{(1)T} J \d V \wedge \d \jo_+^{(2)}> d\l 
          - \R \frac{1}{ a} < \jo_+^{(2)T} J \d V \wedge \d \jo_-^{(1)}> d\l  \label{TT}
\end{eqnarray}
Applying the contraction operator to  the first term in \ref{TT}

\begin{eqnarray}
i_{\partial_t}\,  \R \;\frac{1}{ a}& < &\jo_-^{(1)T} J \d V \wedge \d \jo_+^{(2)}> d\l \nonumber \\
               & & = \R\;  \frac{1}{ a} < \jo_-^{(1)T} J V^{\bullet}  \d \jo_+^{(2)}> d\l
             -   \R \;\frac{1}{ a} < \jo_-^{(1)T} J \d V   \jo_+^{(2)\bullet}> d\l. \nonumber
\end{eqnarray}
From \ref{V} we have 
$$
\frac{1}{ a}= a_0+\frac{a_1}{ \l}+\ldots, \qquad\qquad{\rm where} \qquad a_0=1, \qquad a_1=-i \int_{-\infty}^{+\infty} |\psi|^2.
$$
Using $V^{\bullet}=\[ \frac{i}{2} \sigma_3, V\]=i \sigma_3 V$, and Lemma 3.1, we have 
$$
   \R\; \frac{1}{ a} < \jo_-^{(1)T} J V^{\bullet}  \d \jo_+^{(2)}> d\l
=a_0< \hat \jo_0^T J i \sigma_3 V \d \jo_1>= -<\psi \d \psib>.
$$
Similarly, using $\jo^{\bullet} = \frac{i}{ 2} \sigma_3 \jo$, we have
\begin{eqnarray*}
   \R\; \frac{1}{ a} < \jo_-^{(1)T} J \d V   \jo_+^{(2)\bullet}> d\l 
  & = & a_1<\hat \jo_0^T J \d V \frac{i}{ 2}\sigma_3 \jo_0> \\
 & + &   a_0\[  <\hat \jo_0^T J \d V \frac{i}{ 2}\sigma_3 \jo_1> +
<\hat \jo_1^T J \d V \frac{i}{ 2}\sigma_3 \jo_0> \].
\end{eqnarray*}
The first term vanishes, the second produces
$$
 = \frac{1}{ 2} <\psib \d \psi - \psi \d \psib >.
$$
Finally,
$$
i_{\partial_t}  \R  \frac{1}{ a} < \jo_-^{(1)T} J \d V \wedge \d \jo_+^{(2)}> d\l \nonumber = - \d H_1.
$$
The second term in  formula \ref{TT} can be treated similarly. Therefore, 
$$
i_{\partial_t} \tr\, \R \;  <H_+^* J\delta V \wedge \delta H_+> = -2 \d H_1.
$$
The second term in  formula \ref{SY}  produces the same result. 
The proof is finished.  \qed

\subsection{Action--angle variables}

In response to  infinitesimal deformations of the matrix 
$\tilde{V}= V +  \d V $ the matrix $T(\l)$ 
changes according to the rule: $\tilde{T}(\l)=T(\l) + \d   T(\l)  + \hdots$.
The next result is similar to Lemma 2.3  of the periodic case.
\begin{lem}   The following formula holds
$$
< H_+^+ J\d V H_+>=\[\begin{array}{ccccccc}
                            -\d a &  \overline{b}\d a - a \d \overline{b} \\
                           a\d b - b \d a & - \d a \end{array}\], 
$$
with  averaging  defined as
$$
< H_+^+ J\d V H_+> =\int_{-\infty}^{+\infty}H_+^+(x,\l) J\d V(x)  H_+(x,\l)\; dx. 
$$
\end{lem}

\noi
{\it Proof.}  Let us assume, first, that $\psi$ has compact support.
We denote by $\tilde V,\; \tilde T$ and $\tilde \jo$ deformed matrices
$V, T$ and the Jost solution $\jo$.  We will derive the expression for $<\jo_+^{(2)\,T} J\d V \jo_{+}^{(1)}>$ 
in the left-upper corner. First, we obtain the formula 
\bay\label{FF}
 < \jo^+ J\delta V \jo>  + \;\text{lower order terms} = \left.\jo^+ J \tilde \jo \right|_{-L}^{+L}.
\ey
Indeed, 
\begin{eqnarray*}
\jo^+ \([J\partial_x -J \tilde V]\tilde \jo\)& =& 0,\\
\(\jo^+ [J\partial_x -J  V]\)\tilde \jo& =& 0.
\end{eqnarray*}
Subtracting one identity from another we have 
$$
 \jo^+ J\delta V \jo + \;\text{lower order terms}= \jo^+\(J\partial \tilde \jo\) -\( \jo^+ J \partial\) \tilde \jo .  
$$
Integrating the RHS in x variable we obtain 
$$
\int_{-L}^{+L} \[ \jo^+ J \tilde \jo' + \jo^{+\prime} J \tilde \jo \] \, dx =\left.\jo^+ J \tilde \jo
\right|_{-L}^{+L}.
$$
This implies \ref{FF}. 
Now using formulas for the asymptotics of $\jo^{(2)}_+$ and $\jo^{(1)}_-$  for the RHS of \ref{FF},     we have
\begin{eqnarray*}
\jo^{(2)T}_+ J \tilde \jo^{(1)}_-\left. \right|_{-L}^{+L}&=&  \fr^TJ\[ \tilde a \fl +
\tilde b \fr\]\left. \right|^{+L} - 
\[a \fr^T -  \overline{b} \fl^T \]J \fl \left. \right|_{-L}\\
&=&  \tilde{a} \fr^T J \fl - a \fr^T J \fl= a-\tilde{a}=- \d a +\text{lower order terms}. 
\end{eqnarray*}
Collecting terms of the same order,  we obtain the
result. The  case of a potential with non-compact support
can be considered using   approximation arguments. For other entries the arguments are the same. 
Lemma is proved. \qed

\begin{thm} The following formulas hold 
$$
\omega_0= \frac{1}{ \pi i}  \int\limits_{-\infty}^{+\infty}  
\frac{\d \bar b(\l)\wedge \d b(\l)}{ |a(\l)|^2} \, d\l. 
$$
\end{thm}
\noi 
{\it Proof.} By the Cauchy integral formula 
$$
 \frac{1}{ 2} \tr\;   \R   <H_+^* J\delta V \wedge \delta H_+>= -\frac{1}{ 2\pi i} 
\int\limits_{-\infty}^{+\infty} \tr  <H_+^* J\delta V \wedge \delta H_+>  \, d\l, 
$$
and
$$
 \frac{1}{ 2} \tr\;   \R   <H_-^* J\delta V \wedge \delta H_->= \frac{1}{ 2\pi i} 
\int\limits_{-\infty}^{+\infty} \tr  <H_-^* J\delta V \wedge \delta H_->  \, d\l. 
$$
Taking  sum 
$$
\omega_0= \frac{1}{ 2\pi i} 
\int\limits_{-\infty}^{+\infty} \tr  <H_-^* J\delta V \wedge \delta H_->  \, d\l - 
\frac{1}{ 2\pi i} 
\int\limits_{-\infty}^{+\infty} \tr  <H_+^* J\delta V \wedge \delta H_+>  \, d\l. 
$$
Using \ref{RH}, \ref{ARH}  
$$
\d H_-=\d H_+ S +H_+\d S, 
$$ 
we obtain  
\begin{eqnarray*}
\omega_0& =&   \frac{1}{ 2\pi i}
\int\limits_{-\infty}^{+\infty} \tr\;  S^{-1} <H_+^* J\delta V \wedge \delta H_+> S \;   \, d\l 
   +\int\limits_{-\infty}^{+\infty} \tr\;  S^{-1} <H_+^* J\delta V  H_+> \wedge \d S \,   \, d\l\\
& &-\frac{1}{ 2\pi i}\int\limits_{-\infty}^{+\infty}\tr <H_+^* J\delta V\wedge \delta H_+>  \,d\l\\
&=& \int\limits_{-\infty}^{+\infty} \tr  <H_+^* J\delta V  H_+> \wedge \d S S^{-1}  \, d\l.
\end{eqnarray*}
Now, applying the result of Lemma 3.5, we have
$$
\omega_0=  \frac{1}{ 2\pi i} \int\limits_{-\infty}^{+\infty} \tr\;- \frac{ \sigma_3}{ a^*} 
\[\begin{array}{ccccccc}
                            -\d a &  \overline{b}\d a - a \d \overline{b} \\
                           a\d b - b \d a & - \d a \end{array}\]
\wedge \d S\, S^{-1} \, d\l.
$$
After simple algebra we arrive at the stated identity.  Theorem  is proved. \qed

\noi
{\it Remark 1.}  The formula of Theorem  can be   put easily into more familiar form using the identities
$$
|a|^2-|b|^2=1,\qquad\qquad\qquad \d \log |b|^2=\frac{2|a|\d|a|}{|a|^2-1}.
$$
Indeed,
\bey
  \frac{1}{  i}    \frac{\d \bar b(\l)\wedge \d b(\l)}{ |a(\l)|^2} 
&=& \frac{|b(\l)|^2 \d \log | b(\l)|^2 \wedge \d {\rm ph}\, b(\l)}{ |a(\l)|^2} \\
&=& 2  \d \log|a(\l)|\wedge  \d\, {\rm ph}\, b(\l).
\eey
Therefore, 
\bey
\omega_0=  \frac{1}{ \pi i}  \int\limits_{-\infty}^{+\infty}  
\frac{\d \bar b(\l)\wedge \d b(\l)}{ |a(\l)|^2}= \frac{2}{ \pi} \int\limits_{-\infty}^{+\infty} \d \log|a(\l)|\wedge  \d\, {\rm ph}\, b(\l)  
\;  d\l.
\eey

\noi
{\it Remark 2.}  The formula
$$
\omega_n= \frac{1}{ \pi i}  \int\limits_{-\infty}^{+\infty}  
\frac{\d \bar b(\l)\wedge \d b(\l)}{ |a(\l)|^2} \l^n\, d\l, \qquad\qquad\qquad n=1,2,\ldots;
$$
subject to the constrains $H_k=const,\; k=1,\ldots,n$; gives Darboux coordinates for higher symplectic forms.

\

\end{document}